\documentclass[prd, superscriptaddress, nofootinbib, floatfix]{revtex4-1}

\usepackage{tikz}
\usepackage{xcolor}
\usepackage[subfigure]{graphfig}
\usepackage[normalem]{ulem} 
\usepackage{wrapfig}
\usepackage{amsmath}
\usepackage{amsfonts}
\usepackage{amssymb}
\usepackage{booktabs}
\usepackage{multirow}
\usepackage{slashed}
\usepackage{physics}
\usepackage{tabularx}
\usepackage{soul}
\usepackage{ulem}
\usepackage{hhline}
\usepackage{float}
\usepackage{enumitem}
\usepackage[colorlinks,pdfusetitle]{hyperref}
\hypersetup{colorlinks=true,allcolors=[rgb]{0,0.0,1.0}}
\usepackage{longtable}
\usepackage{relsize}
\usepackage{verbatim}

\usepackage{nccmath}
\usepackage{esvect}

\newcommand\befs{\begin{figure*}[ht!]}
\newcommand\eefs[1]{\label{fig:#1}\end{figure*}}
\newcommand\bef{\begin{figure}[ht!]}
\newcommand\eef[1]{\label{fig:#1}\end{figure}}
\newcommand\beq{\begin{equation}}
\newcommand\eeq[1]{\label{#1}\end{equation}}
\newcommand\beqa{\begin{eqnarray}}
\newcommand\eeqa[1]{\label{#1}\end{eqnarray}}
\newcommand\bet{\begin{table}}
\newcommand\eet[1]{\label{tb:#1}\end{table}}
\newcommand\bets{\begin{table*}}
\newcommand\eets[1]{\label{tb:#1}\end{table*}}

\def\be{\begin{equation}}
\def\ee{\end{equation}}
\newcommand{\bea}{\begin{eqnarray}}
\newcommand{\eea}{\end{eqnarray}}
\newcommand{\ba}{\begin{align}}
\newcommand{\ea}{\end{align}}

\newcommand{\Dl}{\Delta}
\newcommand{\wt}{\widetilde}

\newcommand{\nn}{\nonumber}
\newcommand{\om}{\omega}
\newcommand{\Del}{\Delta}
\newcommand{\ms}{\overline{\rm{MS}}}
\usepackage[ruled,vlined,linesnumbered,noresetcount]{algorithm2e}
\usepackage{graphicx}
\graphicspath{{figures/}}

\begin{document}

\widetext

\title{Polarized and unpolarized gluon PDFs: generative machine learning applications for lattice QCD matrix elements at short distance and large momentum} 

\newcommand{\DUP}{Department of  Physics, University of Dhaka, Dhaka 1000, Bangladesh}\affiliation{\DUP}
\newcommand{\KU}{Department of Physics and Astronomy, University of Kansas, Lawrence, Kansas 66045, USA}\affiliation{\KU}
\newcommand{\BNL}{Physics Department, Brookhaven National Laboratory, Upton, NY 11973, USA}\affiliation{\BNL}
\newcommand{\RBRC}{RIKEN-BNL Research Center, Brookhaven National Laboratory, Upton, NY 11973, USA}\affiliation{\RBRC}
\newcommand{\SNL}{Systems biology, Sandia National Laboratory, Livermore, CA 94550, USA}\affiliation{\SNL}
\newcommand{\APS}{American Physical Society, Hauppauge, New York 11788}\affiliation{\APS}
\newcommand{\DU}{Department of Theoretical Physics, University of Dhaka, Dhaka 1000, Bangladesh}\affiliation{\DU}
\newcommand*{\SDU}{Key Laboratory of Particle Physics and Particle Irradiation (MOE), Institute of Frontier and Interdisciplinary Science, Shandong University, Qingdao, Shandong 266237, China}\affiliation{\SDU}
\newcommand{\UMN}{School of Physics and Astronomy, University of Minnesota, Minneapolis, MN 55455, USA}\affiliation{\UMN} 
\newcommand{\NMSU}{Department of Physics, New Mexico State University, Las Cruces, NM 88003, USA}\affiliation{\NMSU}

\author{Talal~Ahmed~Chowdhury}\affiliation{\DUP}\affiliation{\KU}
\author{Taku~Izubuchi}\affiliation{\BNL}\affiliation{\RBRC}
\author{Methun~Kamruzzaman}\email{mkamruz@sandia.gov}\affiliation{\SNL}
\author{Nikhil~Karthik}\affiliation{\APS}
\author{Tanjib~Khan}\affiliation{\DU}
\author{Tianbo~Liu}\email{liutb@sdu.edu.cn}\affiliation{\SDU}
\author{Arpon~Paul}\affiliation{\UMN}
\author{Jakob~Schoenleber}\affiliation{\BNL}\affiliation{\RBRC}
\author{Raza~Sabbir~Sufian}\email{gluon2025@gmail.com}
\affiliation{\NMSU}\affiliation{\BNL}\affiliation{\RBRC}

\begin{abstract}
Lattice quantum chromodynamics (QCD) calculations share a defining challenge by requiring a small finite range of spatial separation $z$ between quark/gluon bilinears for controllable power corrections in the perturbative QCD factorization, and a large hadron boost $p_z$ for a successful determination of collinear parton distribution functions (PDFs). However, these two requirements make the determination of PDFs from lattice data very challenging. We present the application of generative machine learning algorithms to estimate the polarized and unpolarized gluon correlation functions utilizing short-distance data and extending the correlation up to $zp_z \lesssim 14$, surpassing the current capabilities of lattice QCD calculations. We train physics-informed machine learning algorithms to learn from the short-distance correlation at $z\lesssim 0.36$ fm and take the limit, $p_z \to \infty$, thereby minimizing possible contamination from the higher-twist effects for a successful reconstruction of the polarized gluon PDF. We also expose the bias and problems with underestimating uncertainties associated with the use of model-dependent and overly constrained functional forms, such as $x^\alpha(1-x)^\beta$ and its variants to extract PDFs from the lattice data.  We propose the use of generative machine learning algorithms to mitigate these issues and present our determination of the polarized and unpolarized gluon PDFs in the nucleon. 
\end{abstract}

\maketitle


\section{Introduction} \label{sec:intro1}

The determination of parton distribution functions (PDFs) from lattice quantum chromodynamics is of particular theoretical interest to explore the non-perturbative sector of QCD from the first principles. Precise and accurate knowledge of the universal nonperturbative PDFs sheds light on our understanding of the structure of hadrons in terms of quarks and gluons, the fundamental degrees of freedom in QCD. In particular, gluons, which serve as mediator bosons of the strong interaction, play a key role in the nucleon’s mass and spin structures. Accurate knowledge of PDFs is also essential for analyzing and interpreting physics from various scattering experiments. 


To achieve the goal of calculating momentum fraction, $x$-dependent nonperturbative structure functions and PDFs from the first-principles lattice QCD (LQCD) calculations, there have been several proposals, such as the path-integral formulation of the deep-inelastic scattering hadronic tensor~\cite{Liu:1993cv,Liu:1999ak}, the operator product expansion~\cite{Detmold:2005gg}, current-current correlations~\cite{Braun:2007wv}, the Compton amplitude approach~\cite{QCDSF:2012mkm,Chambers:2017dov}, quasi-PDFs and large momentum effective theory (LaMET)~\cite{Ji:2013dva, Ji:2014gla},  lattice cross-sections~\cite{Ma:2014jla,Ma:2017pxb,Sufian:2019bol}, and pseudo-PDFs~\cite{Radyushkin:2017cyf}. While LQCD calculations have made significant computational achievements in calculating PDFs (see the recent reviews~\cite{Ji:2020ect,Constantinou:2022yye} and the references therein), one defining challenge is that the bilocal light-cone correlators that are necessary to determine the PDFs cannot be evaluated directly on the Euclidean lattice. Quasi-PDFs framework~\cite{Ji:2013dva} circumvents this drawback by calculating equal-time Euclidean nonlocal matrix elements with hadron states at non-zero momentum, $p_z$. The corresponding quasi-PDFs can be matched to the light-cone PDFs when the hadron momentum is large, by applying the LaMET formalism~\cite{Ji:2014gla}. On the other hand, the pseudo-PDFs framework uses short-distance factorization and can be perturbatively matched to the light-cone PDF. To implement the QCD short-distance factorization with controllable power corrections and because the renormalons in the renormalization of the  LQCD matrix elements at large distances become significantly important~\cite{Braun:2018brg}, it is desirable to use only the short-distance matrix elements. However, one needs LQCD matrix elements at large Ioffe-time $\om \equiv -z\cdot p$~\cite{Gribov:1965hf,Ioffe:1969kf,Braun:1994jq} while keeping $z$ small and making the hadron boost $p_z$ along the $z$-direction as large as possible to facilitate the inverse problem in determining PDFs from a limited $\om$-range data. The problem of using large-$z$ matrix elements has been found in a recent calculation~\cite{Bhat:2022zrw} with the implementation of the 2-loop matching~\cite{Li:2020xml} as the higher-twist contribution can become significant for $z\gtrsim 0.35$ fm  (see for other findings in~\cite{Karthik:2021sbj,Ji:2022ezo,Su:2022fiu}). An alternative proposal is to use the pseudo-PDFs ratio scheme~\cite{Radyushkin:2017cyf} only at small $z$ in hybrid-  and self-renormalization schemes~\cite{Ji:2020brr,LatticePartonCollaborationLPC:2021xdx}, treating short and long-distance scales separately as well as using model-dependent extrapolations of lattice correlation functions at large $\om$ to facilitate the determination of PDFs within the LaMET framework.

In this work, we train the generative machine learning (ML) algorithms using short-distance LQCD correlation functions and focus on the generation of the LQCD matrix elements at large $\om$ which can be used to determine the polarized gluon $x\Dl g(x)$ and the unpolarized gluon $xg(x)$ PDFs. The principle behind training ML algorithms with short-distance correlation functions is based on the ability of the ML to learn effectively from the data points where higher-twist contamination is minimal. Therefore, it is expected that the generation of the renormalized matrix elements at large $\om$ will also involve small contamination from higher-twist effects. Moreover, the ML algorithms predicting the LQCD matrix element at large $\om$ for the larger values of $p_z$ will not only be extremely useful for handling the inverse problem of reconstructing the PDFs but also be particularly helpful in future calculation of quark and gluon PDFs within the LaMET fromalism~\cite{Ji:2014gla}.

We note that there have been a few calculations of the unpolarized gluon PDFs in the nucleon, pion, and kaon~\cite{Fan:2020cpa,Fan:2021bcr,Fan:2022kcb,Salas-Chavira:2021wui,HadStruc:2021wmh,Delmar:2023agv,Good:2023ecp,Good:2023rbp,Good:2024iur}, as well as matrix element calculations toward the gluon helicity Ioffe-time distribution~\cite{HadStruc:2022yaw}, and the first LQCD determination of the gluon helicity PDF~\cite{Khan:2022vot}.  In our work,  we will show that determining the light-cone quark and gluon helicity PDFs require LQCD data in the limit of $p_z \to \infty$. We will demonstrate how physics-informed ML can help attain this limit and enable us to determine helicity PDFs, particularly the gluon helicity PDF. We will use published LQCD matrix elements of the unpolarized and polarized gluon correlation functions~\cite{Balitsky:2019krf, Balitsky:2021cwr} from Refs.~\cite{HadStruc:2021wmh,HadStruc:2022yaw,Khan:2022vot}. The essential goals of this work are as follows:
\begin{itemize}

\item Application of physics-informed ML algorithms for constructing PDFs at larger momentum. 

\item Highlight the bias and underestimation of uncertainties associated with model-dependent extraction of PDFs.

\item Removal of a contamination term from the Euclidean matrix elements that hinders the extraction of $x\Dl g(x)$ from LQCD data, and train ML algorithms using short-distance data to expose and minimize potential contamination from the higher-twist effects.

\item Neural network determination of the polarized and the unpolarized gluon PDFs using ML-generated data at large $\om$ while making an effort to avoid underestimating uncertainties.
\end{itemize}
The paper is organized as follows. In Sec.~\ref{sec:extrapol}, we first discuss the necessity of extrapolating LQCD data outside the available range of $\om$ and discuss how this extrapolation has been performed previously. This section reveals the risk of using model-dependent and overly constrained PDF ansatz and motivates the need for ML to generate data at large $\om$ where LQCD data are not available. In Sec.~\ref{sec:whyML}, we discuss the theoretical essence of LQCD data at large momentum and the prospect of generative ML models to overcome the above-mentioned problems. We discuss a prescription to eliminate the $p_z$-dependence in the lattice data of quark and gluon helicity correlation functions. In Secs.~\ref{sec:ML1} and~\ref{sec:MLdisc}, we explore three different generative ML algorithms to determine the polarized and unpolarized gluon Ioffe-time distributions~\cite{Braun:1994jq} beyond the reach of present-day LQCD calculations. In Secs.~\ref{sec:PDFs} and~\ref{sec:PDFdisc}, we determine $x\Dl g(x)$ and $xg(x)$ distributions using neural network on the ML-generated Ioffe time distributions and highlight some of the important implications of this work. Sec.~\ref{sec:con} contains our concluding remarks and outlook.

\section{Necessity of LQCD data at large $\om$, higher-twist contaminations, and biases in ansatz-depedent PDFs extraction  }\label{sec:extrapol}

In general, LQCD calculation of matrix elements at hadron momentum $p_z > 2~{\rm to}~3$ GeV becomes exponentially challenging due to poor signal-to-noise ratio. Specifically, achieving precise results for the gluonic matrix elements at a lighter pion mass and $p_z > 2$ GeV will be very challenging in the near future LQCD calculations as it involves Wick contractions associated with the lattice calculations of disconnected insertions. The state-of-art LQCD calculations of the unpolarized gluonic matrix elements or the Ioffe-time distribution (ITD) have provided data up to $\om_{\rm max}\approx 7$~\cite{HadStruc:2021wmh,Fan:2022kcb} or even smaller~\cite{Delmar:2023agv}, beyond which the data is consistent with zero due to large noise. One way to facilitate the extraction of PDFs from the lattice data in a wide range of $\om$,  is to perform LQCD calculation at large spacelike separations at a fixed $p_z$. However, this poses the risk of uncontrollable power corrections, and the factorization of the LQCD matrix elements into nonpertubative PDF and perturbative coefficients might not be sensible. Additionally, at a fixed $p_z$, lattice data becomes much noisier as $z$ increases. Therefore, when one parameterizes the PDFs using some model-constrained functional forms and includes possible higher-twist contributions such as $\mathcal{O}(\Lambda_{\rm QCD}^2z^2)$ or some higher order $\mathcal{O}(z^n)$ terms in parameterizing the data, the noisy data may not allow for proper isolation of the higher-twist contributions from the leading twist-2 contribution (for example, see Fig.~5 in~\cite{HadStruc:2022nay} at $z=0.75$ fm and $p_z \geq 1.67$ GeV). 

In addition, constrained by a limited range of $\om$, the available lattice ITDs are inherently sensitive to only the first few moments, as shown in~\cite{Mankiewicz:1996ep} and reiterated in~\cite{Gao:2020ito,Sufian:2020wcv,Su:2022fiu}. Lattice calculations using model-dependent parametrization of PDFs utilize the correlation between the first two or three accessible moments to model the missing information beyond the available $\om_{\rm max}$ and attempt to extract PDF in the full $ x$ range. In addition, for currently available LQCD calculations in a limited $\om$ range, these functional forms can be biased, leading to an unreliable $\chi^2/{\rm d.o.f.}$, and underestimation of uncertainties. For example, the parametrization $x^\alpha(1-x)^\beta$ for the LQCD determination of $xg(x)$ distribution led to a diverging PDF in~\cite{Fan:2020cpa} and a converging PDF in~\cite{HadStruc:2021wmh}  at small $x$. However,  none of these lattice ITDs reach the Regge region or have much sensitivity to the small-$x$ physics~\cite{Sufian:2020wcv}. In~\cite{HadStruc:2021wmh}, $\alpha \geq 0$ constraint was imposed in a Bayesian fit, motivated by a phenomenological analysis in~\cite{Sufian:2020wcv}. Otherwise, it would have resulted in a diverging PDF as in~\cite{Fan:2020cpa}. A similar observation can be made for $xg(x)$ extractions in recent lattice calculations~\cite{Fan:2018dxu,Fan:2021bcr,Fan:2022kcb,Salas-Chavira:2021wui,Delmar:2023agv,Good:2023ecp} where all these lattice ITDs are limited in ITD the range of $\om \lesssim 6$. It is therefore desirable to determine LQCD matrix elements at large $\om$, while maintaining controllable power corrections. Additionally, it is important not to rely on simple ansatz fits of PDFs using data in a limited $\om$ range as the resulting PDFs can depend heavily on the chosen functional forms, priors, and other strong constraints used in the fits. None of the existing LQCD data is sensitive to either very large $x$ or very small $x$ PDFs.  It is not evident why a product of $x^\alpha$ (governed by the Regge behavior~\cite{Regge:1959mz}) and $(1-x)^\beta$ (governed by the perturbative QCD counting rule~\cite{Brodsky:1973kr,Lepage:1980fj,Brodsky:1994kg}) should describe PDFs in the mid-$x$ region, where the lattice data is more sensitive to. 

As an illustration of the problem, by reconstructing the unpolarized gluon ITD from NNPDF distribution~\cite{NNPDF:2017mvq}, it can be shown that the NNPDF ITD has significantly smaller uncertainty across the entire $\om$ range compared to the lattice ITD presented in~\cite{HadStruc:2021wmh}. However, the large-$x$ PDF determined from the LQCD data has much smaller uncertainty than the NNPDF distribution. Although the Fourier transform of $\tilde{\om} = \frac{1}{x}$~\cite{Ma:2017pxb,Miller:2019ysh} and precise lattice data at small $\om$ can be helpful to constrain PDF in the large-$x$ region, such a dramatic reduction of the uncertainty in the large-$x$ PDFs is not expected. This possibility of underestimating the uncertainty in the lattice calculation of PDFs has been elaborated in Fig.~\ref{fig:lattITD}, where the lattice and the phenomenological ITDs $\mathcal{I}_g(\om,\mu)$ are chosen to be at $\mu=2$ GeV matching scale in the $\ms$ scheme. Keeping in mind that this is not a one-to-one comparison because of the different types of observables used in the determination of PDFs, the reconstructed PDF from the lattice data using a Bayesian fit with the functional form of PDF $x^\alpha(1-x)^\beta$ (with imposed constraint $\alpha \geq 0$) in Fig.~\ref{fig:lattITD} has uncertainty smaller than the NNPDF determination in the whole $x \geq 0.7$ region. Therefore, it can be misleading to obtain a much more precise PDF from noisier lattice data in a limited $\om$ range. The reason for this unrealistic precision in~\cite{HadStruc:2021wmh} stems from two constraints:
\begin{itemize}
    \item The $(1-x)^\beta$ term in the PDF functional form shrinks the uncertainty in the moderate to large $x\to 1$ region.
    \item Bayesian fits using Jacobi polynomial basis~\cite{AtashbarTehrani:2007odq,Khorramian:2009asi,Taghavi-Shahri:2016idw} with a constraint $\alpha \geq 0$.
\end{itemize} 

A more direct comparison using the same lattice data with or without using constrained functional forms in the determination of gluon helicity PDFs has been illustrated in the lower panel in Fig.~\ref{fig:lattITD}~(from Ref.~\cite{Khan:2022vot}) which shows the severity of underestimating uncertainties when constrained PDF ansatz is used.

In Sec.~\ref{sec:PDFs}, we will determine the polarized and unpolarized gluon PDFs using ML-generated data at large $\om$ to alleviate the inverse problem. To avoid  the underestimation of uncertainty in the lattice determination of PDFs as discussed above, we will resort to the neural network analysis in the subsequent part of the manuscript. We acknowledge that due to the limited lattice data, the neural network may not explore the full spectrum of the potential architectures and solutions. However, it should strive for a level of generality beyond specific parameterizations.

\befs 
\centering

\includegraphics[scale=0.6]{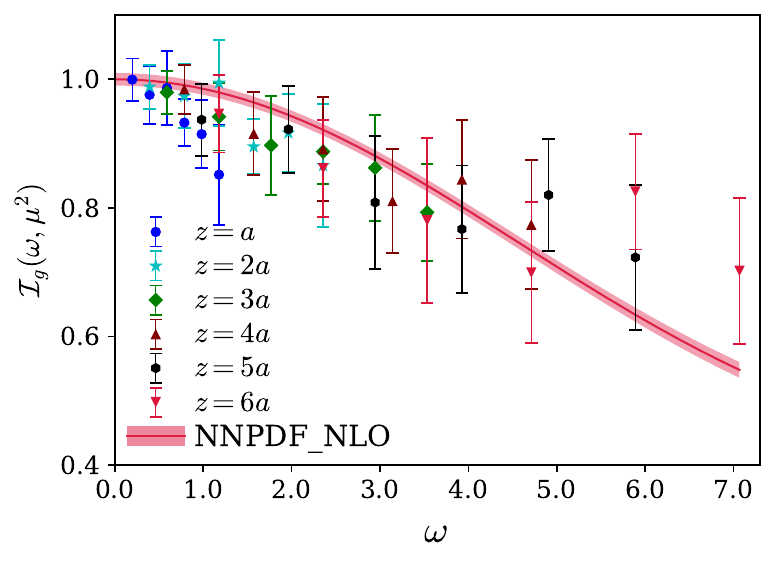}
\includegraphics[scale=0.6]{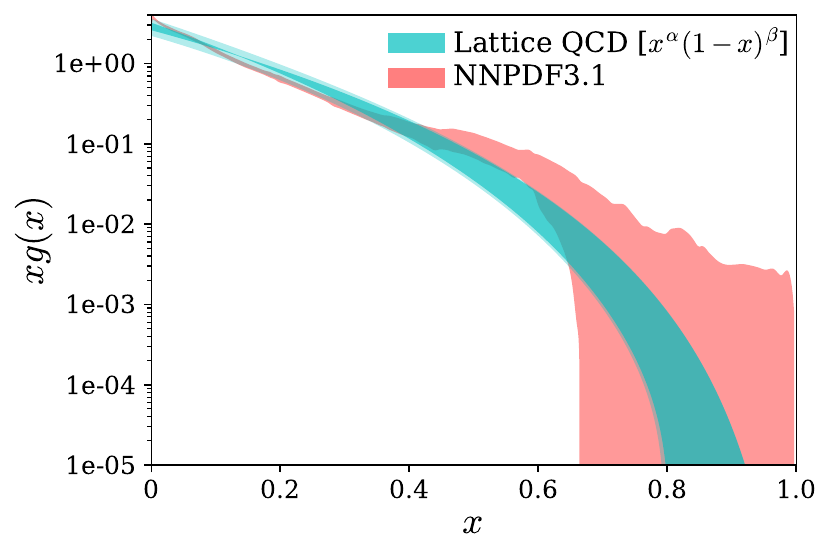}
\includegraphics[scale=0.6]{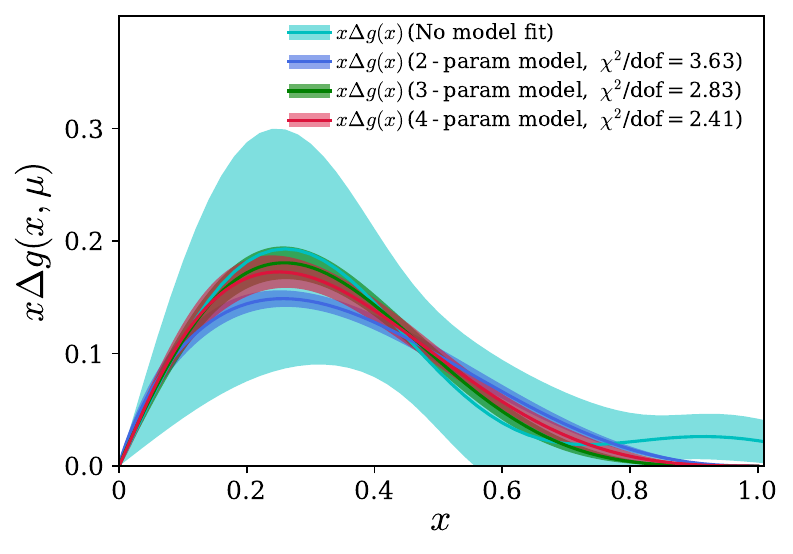}

\caption{\label{fig:lattITD} 
Upper left panel: Ioffe-time distribution after the implementation of the perturbative matching kernel on the lattice reduced pseudo-ITD calculated for a $2$-parameter model fit, in the $\ms$ renormalization scheme at 2 GeV in Ref.~\cite{HadStruc:2021wmh}. The $z=a-6a$ denotes the length of the field separation as a function of the lattice spacing $a$. The Ioffe-time distribution of the NNPDF distribution is normalized to $1$ at $\om=0$ for a comparison with the LQCD matrix elements. Upper right panel: unpolarized gluon PDF (cyan band) extracted from the lattice data in Ref.~\cite{HadStruc:2021wmh} using the $2$-parameter model fit $x^\alpha(1-x)^\beta$. We compare our results to the gluon PDF extracted from the NNPDF3.1 global fit~\cite{Ball:2017nwa}. Lower panel: a similar example is given for the gluon helicity PDF from~\cite{Khan:2022vot}. The cyan band is determined directly from the lattice QCD data without assuming any functional form for the PDF ansatz, while the other bands with smaller uncertainties result from different model fits used to extract the $x\Dl g(x)$ distribution. The 2-parameter fit is done using the fit-expression $x^\alpha (1 - x)^\beta$, and subsequently by introducing additional parameters for 3 and 4 parameter fits, namely $\rho$ and $\gamma$, resulting
in the fit-expression $x^\alpha(1 - x)^\beta (1 + \rho \sqrt{x} + \gamma x)$. }     
\eefs{mockdemocn}

\section{Theoretical essence of large momentum data and opportunity of generative machine learning applications}\label{sec:whyML}

Building on the findings presented in~\cite{Khan:2022vot}, this section briefly outlines the prescription for removing contamination terms in the Euclidean matrix elements that may obstruct the extraction of light-cone quark and gluon helicity PDFs. 

To determine $\Dl g(x)$, we calculate matrix elements of the gluon field strength tensor $G_{\mu\nu}$ and its dual $\widetilde{G}_{\lambda\beta}=(1/2)\epsilon_{\lambda\beta\rho\gamma}G^{\rho\gamma}$ separated by a spatial Wilson line $W[z, 0]$~\cite{Ji:2013dva,Balitsky:2021cwr},
\begin{eqnarray}\label{eq1:ME}
\Dl M_{\mu\alpha;\lambda\beta}(z,p,s) = \bra{p,s}G_{\mu \alpha} (z) \, W[z, 0] \widetilde{G}_{\lambda \beta} (0) \ket{p,s}
 - (z \to -z),
\end{eqnarray}
where $z$ is the separation between the gluon fields, $p$ is the nucleon four-momentum, and $s$ is the nucleon polarization. The combination that can be used to determine the gluon helicity correlation with the least number of contamination terms is~\cite{Balitsky:2021cwr} 
\bea \label{eq:lattmat}
\Dl {\mathcal M}_{00}(z,p_z)\equiv \Dl M_{0i;0i}(z,p_z) + \Dl M_{ij;ij}(z,p_z),
\eea
$i,j=x,y$ being perpendicular to the nucleon boost in the $z$-direction, $p= \{p_0, 0_\perp, p_z \}$. Utilizing the multiplicative renormalizability of the quasi-PDF matrix
elements~\cite{Izubuchi:2018srq,Ji:2017oey,Zhang:2018diq,Li:2018tpe}, one can use the pseudo-PDFs ratio method~\cite{Radyushkin:2017cyf} and obtain the renormalization group invariant reduced pseudo-ITD, $\Dl\mathfrak{{M}}$~\cite{Balitsky:2021cwr}:
\bea \label{eq:rITDdef}
\Dl\mathfrak{{M}}(z,p_z)\equiv i \frac{[\Dl{\mathcal M}_{00}(z,p_z)/p_z { p_0}]/Z_{\rm L}(z/a_L)}{{\cal M}_{00}(z,p_z=0)/m_p^2}\, ,
\eea
where, $m_p$ is the nucleon mass, ${\cal M}_{00}(z,p_z) \equiv [M_{0i;i0}(z,p_z)+M_{ji;ij}(z,p_z)]$ is the spin-averaged matrix element related to the unpolarized gluon correlation~\cite{Balitsky:2019krf,HadStruc:2021wmh} and the factor $1/Z_{\rm L} (z/a_L)$ is determined in~\cite{Balitsky:2021cwr} to cancel the UV logarithmic vertex anomalous dimension of  $\Dl {\cal M}_{00}$. 

$\Dl\mathfrak{{M}}$ can be expressed in terms of invariant amplitudes, $\Dl\mathcal{M}_{sp}^{(+)}$ and $\Dl{\mathcal M}_{pp}$~\cite{Balitsky:2021cwr},
\begin{eqnarray} \label{eq:Ipform1}
\Dl \mathfrak{M}(\om,z^2) =  [\Dl {\mathcal M}_{sp}^{(+)}(\om,z^2) - \om \Dl{\mathcal M}_{pp}(\om,z^2) ]
 - \frac{m_p^2}{p_z^2} \om \Dl{\mathcal{M}}_{pp}(\om,z^2) \, .
\end{eqnarray}
In contrast, the light cone correlation that gives access to $x\Dl g(x,\mu)$ at a scale $\mu$ is
\bea \label{eq:lcITD}
\Dl{\mathcal{I}}_g(\om,\mu) \equiv i [\Dl{\mathcal{M}}_{sp, lc}^{(+)}(\om,\mu) - \om \Dl{\mathcal{M}}_{pp, lc}(\om,\mu)] 
=\frac{i}{2} \int_{-1}^{1} d x~ e^{-ix\om}x\Dl g(x,\mu),
\eea
where the subscript $lc$ denotes taking the light-cone limit $z = z_-$ and applying the UV renormalization resulting in the dependence on the factorization scale $\mu$. Note that light-cone ITD in Eq.~\eqref{eq:lcITD} does not contain the $m_p^2/p_z^2$ suppressed term as appeared in Eq.~\eqref{eq:Ipform1}. An alternative expression of $\Dl \mathcal{M}_{00}(z,p_z)$ shows that this matrix element is nonvanishing at $p_z=0$ and one can remove the $\mathcal{O}(\om)$ contamination following the method describing in~\cite{HadStruc:2022yaw}. However, the residual higher-order contamination can become significant at large $\om$. In order to eliminate the $(m_p^2/p_z^2)\om \Dl{\mathcal{M}}_{pp}$ contribution, we multiply Eq.~\eqref{eq:Ipform1} by  the corresponding lattice squared-momentum $p_k^2$ and obtain~\cite{Khan:2022vot}

\begin{equation} 
p_k^2\Dl \mathfrak{M}(\om)\big\vert_{p_k} =  p_k^2[\Dl {\mathcal M}_{sp}^{(+)}(\om) - \om \Dl{\mathcal M}_{pp}(\om) ] |_{p_k}
 - m_p^2 \om \Dl{\mathcal{M}}_{pp}(\om) |_{p_k}\, .
 \label{eq:momentum}
\end{equation}
Here we have dropped the $z^2$ argument since $z^2$ is determined by specifying $\omega$ and $p_z$. One can then eliminate the $m_p^2 \om \Dl{\mathcal{M}}_{pp}$ contributions across different $p_z$-matrix elements and determine 
\bea
\label{eq:master}
\Dl \mathfrak{M}_g(\om) \equiv \frac{r^2 \Dl \mathfrak{M}(\om)\big\vert_{p_k}-\Dl \mathfrak{M}(\om)\big\vert_{p_l}}{r^2-1} \approx \Dl {\mathcal M}_{sp}^{(+)}(\om) - \om \Dl{\mathcal M}_{pp}(\om),
\eea
which is free of the contamination term and can be matched to light-cone ITD $\Dl{\mathcal{I}}_g(\om,\mu)$. In Eq.~\eqref{eq:master}, different lattice boosts $p_k$ and $p_l$ are related by the ratio $r=p_k/p_l=k/l$ ($k>l$). Note that on the right-hand-side of Eq.~\eqref{eq:master}, we have  not specified the dependence $\Dl {\mathcal M}_{sp}^{(+)}$ and $\Dl{\mathcal M}_{pp}$ on $z^2$. The subtraction at different lattice momenta effectively cancels some of the higher twist $\mathcal{O}(z^2)$ contributions. The dependence of the right-hand-side of Eq.~\eqref{eq:master} on $z^2$ is then two-fold. On the one hand, at fixed order in $\alpha_s$ we have logarithmic $z^2$ dependence in $\Dl {\mathcal M}_{sp}^{(+)}$ and $\Dl{\mathcal M}_{pp}$ that arise when they are matched to the light-cone distributions. On the other hand, we have further power corrections, denoted by using the $\approx$ symbol. We therefore expect that the right-hand-side of Eq.~\eqref{eq:master} should depend weakly on $z^2$. Conversely, if Eq.~\eqref{eq:master} does not hold to a good approximation (i.e. if there is a strong $z^2$ dependence) we can conclude that higher twist effects are substantial. This rationale will be used later in the ML algorithm.

As shown in Fig.~\ref{fig:lattdaat} of Sec.~\ref{sec:ML1}, the LQCD matrix elements for the two largest momenta used in this calculation (ranging from $2.05$ to $2.46$ GeV) begin to overlap within uncertainties. With more precise future data at larger momenta, as well as a clearer $z^2$ dependence, ML algorithms can be employed to isolate $m_p^2 \om \Dl{\mathcal{M}}_{pp}(\om)$ for different values of $z$ in Eq.~\eqref{eq:momentum}, without relying on the contamination term removal method described in Eq.~\eqref{eq:master}.

 After removing the contamination term, $\Dl \mathfrak{M}_g$ can be matched to $\Dl{\mathcal{I}}_g$ and one can  calculate the gluon helicity $\Delta G(\mu)$ in the nucleon from the ITD as
\bea
\Delta G(\mu) = \int_0^\infty d\om ~\Dl{\mathcal{I}}_g(\om,\mu). 
\eea
Following the derivation in~\cite{Saalfeld:1997uv}, it can be shown that in a fixed gauge the nonlocal spatial matrix element in Eq.~\eqref{eq:lattmat} reduces to $\Delta G \rightarrow \left(\vec{E}\times \vec{A_\perp} \right)^3$ in the infinite momentum frame, consistent with the findings in~\cite{Ji:2013fga,Hatta:2013gta}, where $\vec{A}_\perp$ is the transverse part of the gauge potential.

As another example, a contamination term is also present in the LQCD matrix elements associated with the quark helicity PDF~\cite{HadStruc:2022nay}. Following the same notations as in~\cite{HadStruc:2022nay}, the Euclidean matrix elements needed for the calculation of quark helicity PDF can be written as
\bea
M_{\mu5}\left(p,z\right)=\bra{N\left(p,\lambda\right)}
\overline{\psi}\left(z\right)\gamma_\mu\gamma_5W
\left(z,0\right)\psi\left(0\right)\ket{N\left(p,\lambda\right)},
\eea
where $\lambda$ denotes the nucleon helicity. For $\mu=3$ (the $z$ direction), this matrix element can be written in terms of  the Lorentz invariant amplitudes:
\bea \label{eq:tildeY}
\wt{\mathcal{Y}}(\om)\big\vert_{p_{k}} =
 \mathcal{Y}(\om)\big\vert_{p_{k}}
+m_p^2 z^2 \mathcal{R}(\om,z^2)
=\mathcal{Y}(\om)\big\vert_{p_{k}}
+m_p^2 \frac{\om^2}{p_k^2} \mathcal{R}(\om)\big\vert_{p_{k}}.
\eea
Both amplitudes $\mathcal{Y}(\om,z^2)$ and $\mathcal{R}(\om,z^2)$ contain leading twist-2 and higher twist contributions. It is the twist-2 contribution contained in the amplitude $\mathcal{Y}(\om,z^2)$ that can be matched to the light-cone quark helicity PDF with controllable power corrections. The matching coefficient from $\mathcal{R}$ does not contribute to the collinear $\ln z^2$-dependence  of the light-cone PDF and one can remove the $\mathcal{R}$-contamination following the prescription above and obtain:
\bea
\label{eq:qh}
\mathcal{Y}(\om) \equiv  \frac{r^2 \wt{\mathcal{Y}}(\om)\big\vert_{p_k}-\wt{\mathcal{Y}}(\om)\big\vert_{p_l}}{r^2-1} .
\eea

Now, the essence of the ML algorithms to utilize the relation in Eq.~\eqref{eq:master} can be understood as the following. The immediate challenge of implementing Eq.~\eqref{eq:master} is that the amplitudes $\Dl{\mathcal M}_{sp}^{(+)}$ and $\Dl{\mathcal M}_{pp}$ are nonperturbative functions and their functional forms are unknown. It has been shown in~\cite{Khan:2022vot} that in the absence of a theoretically known functional form of these nonperturbative amplitudes, the moments expansion fails to describe the data, and one needs to resort to ML to circumvent the ignorance of the functional forms of these amplitudes. Therefore, to parameterize these unknown nonperturbative amplitudes, one can use some ML approaches constrained by Eq.~\eqref{eq:master} which we refer to as the ``physics-informed" ML. Note that, as long as the equality in Eq.~\eqref{eq:master} holds to a good approximation, the lattice correlation functions will be dominated by the leading-twist contribution and can be used to extract the light-cone $x\Dl g(x)$. As a remarkable outcome, we will see in Sec.~\ref{sec:MLMethods} that this relation is beneficial in exposing possible higher-twist contaminations in the gluon correlation functions at $z \gtrsim 0.36$ fm. Therefore, if ML approaches can be developed and trained on the lattice matrix elements at small $z$, where higher-twist contamination is minimal, and then used to generate the ITD at large $\om$, the resulting extrapolated data may exhibit reduced higher-twist contamination even at larger values of $\om$ for a fixed $p_z$.  The ideal scenario is when ample and precise lattice data points are available, allowing ML to determine the correlations between different $z$ and $p_z$ datasets and estimate higher-twist contaminations. However, the gluon matrix elements do not allow us to isolate the $z$-dependence across different $p_z$ datasets. For example, due to the large uncertainties in the data points, we were unable to isolate the $p_z$-dependence in the unpolarized gluon matrix elements.This will be a subject of future study with more precise LQCD matrix elements. In this study, for the noisy gluonic matrix elements, $\om$ is the only input feature in the parametrization of the ML.  This does not imply that different $p_z$ datasets are independent; rather, it indicates that we are unable to separately disentangle the $z$- and $p_z$-dependence from the existing datasets.  This paper aims to lay the groundwork for ML applications on future precise lattice data. Here, we examine how the LQCD data can be extrapolated to an $\om$-region beyond the range of the currently available data, enabling more accurate determination of PDFs while minimizing unwanted contamination using the generative models of ML.

\section{Application of ML algorithms to the lattice data}\label{sec:ML1}

Gluonic matrix elements become considerably noisier than the quark matrix elements at large hadron boosts and large $z$, limiting access to large $\om$. In this context, the application of ML becomes crucial, as it can effectively learn from the small-$z$ lattice data where the higher-twist contamination is minimal. ML can then predict data at large $\om$, enhancing the accuracy of the PDF determination. The framework presented here employs physics-informed ML with architectures specifically designed to adhere to certain theoretical physics constraints. By restricting predictions to those permitted by the theoretical input, this approach reduces modeling errors, ensures a consistent treatment of data points, and accelerates the training process.

As we will demonstrate in the following, due to the small number of observations and limited data in the $\om \in [0,7]$ region, the pattern of the curves outside the available lattice data, along with the constraint in Eq.~\eqref{eq:master} across different $p_z$ datasets, suggests that the ML model fits can benefit from the phenomenological polarized gluon ITD at large $\omega$. In particular, the term ``physics-informed ML'' refers to the combination of the constraint in Eq.~\eqref{eq:master}, which we enforce across different $p_z$ datasets with minimal power corrections, in addition to the contamination term $(m_p^2/p_z^2) \omega \Delta \mathcal{M}_{pp}$ and the phenomenological ITD at large $\omega$.  We note that no such constraints will be applied when fitting the unpolarized gluon ITD data. Therefore, by physics-informed ML, we exclusively refer to the case of the polarized gluon distribution.

In the following, we present analyses using three different ML algorithms aimed at removing the contamination term discussed in Sec.~\ref{sec:whyML} and generating data at large $\om$ for the polarized gluon ITD. Additionally, we provide a similar analysis for determining the unpolarized gluon ITD.

Machine learning is a data-driven algorithm that uses different statistical processes to extract implicit patterns from high dimensional high volume data. The pattern is used for inference or association with unseen data. Based on the data demography, ML algorithms are divided into three categories: a) supervised learning, b) unsupervised learning, and c) reinforcement learning. Supervised learning algorithms are used for the data where each observation is labeled. The unsupervised learning algorithms are used for the data where the label for each observation is missing. Reinforcement learning gathers data from its environment, processes it, and generates the next action. For instance, a drone collects images from its surroundings, processes them, and finds a path to move forward, backward, and upward or downward. In labeled data, the label can be categorical or real-valued. The ML algorithm used for categorical labeled data is called the classification algorithm, and the algorithm used for real-valued labeled data is called the regression algorithm.

\befs 
\centering
\includegraphics[keepaspectratio, width=0.7\textwidth]{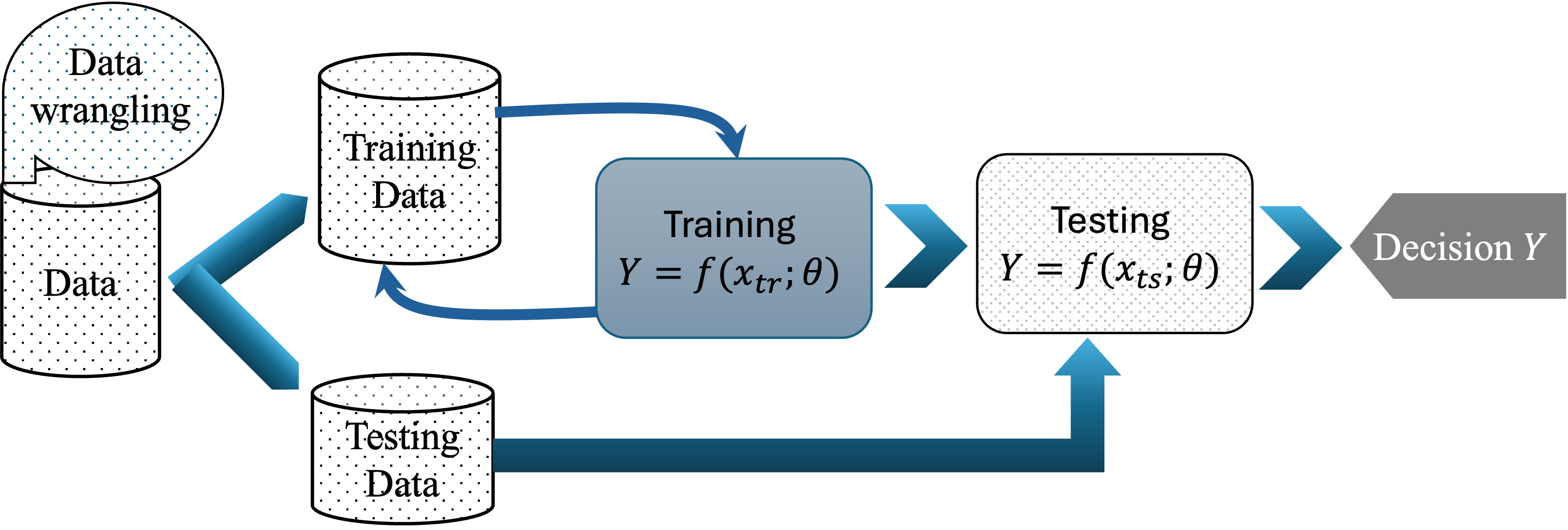}
\caption{The figure shows the necessary components for the machine learning process. The process starts with data. The feature engineering process extracts important features and prepares data for the next level of operations. The data is split into two disjoint subsets: training and testing data. The training data ($x_{tr}$) is used to train a machine learning model, and the testing data ($x_{ts}$) is used to evaluate the performance of the trained model.}     
\eefs{ml-details}

Fig.~\ref{fig:ml-details} shows the general pipeline of any ML algorithm. Data is the most essential element of any ML process. Most of the time, the collected data needs pre-processing, which is called data-wrangling. In the data-wrangling process, one might remove missing and inconsistent data. This step also includes feature engineering, which aims to consider only the important features and data transformation. One good practice is normalizing the features' values to make them comparable. The preprocessed data is split into two disjoint subsets: training and testing data. Based on the volume of data, training data is further subdivided into two disjoint subsets named training and validation. The split proportion is user-defined, but the general convention is to $80\%$ training and $20\%$ testing (or $10\%$ testing and $10\%$ validation).

The training data is used to train an ML model. A model is a function of input variables $X$ and unknown variables $\theta$. It is referred to as parameter because the model learns the values of $\theta$ from data. Another set of variables is required to manipulate the ML system called hyperparameter, for instance, learning rate. These kinds of variables are like settings of an ML algorithm while training. Once the model is trained, we test data with the same parameter and hyperparameter configuration. 


For instance, a decision tree (DT)  is a rule-based machine learning algorithm where it divides the data in a tree fashion but visually represents it in a reverse way. It uses \emph{Gini index}~\cite{daniya2020classification} to choose the best feature to split the data. The root node is a feature that has the highest Gini index, and each of the left nodes holds a decision. As the tree grows, it holds all the information of the training data and may fail to classify some instances for testing data. To avoid this overfitting scenario, a hyperparameter named ``maxheight" controls the tree's growth. Like all hyperparameters, the user can control this variable and prune the tree to control overfitting.

\befs 
\centering
\includegraphics[keepaspectratio, width=0.7\textwidth]{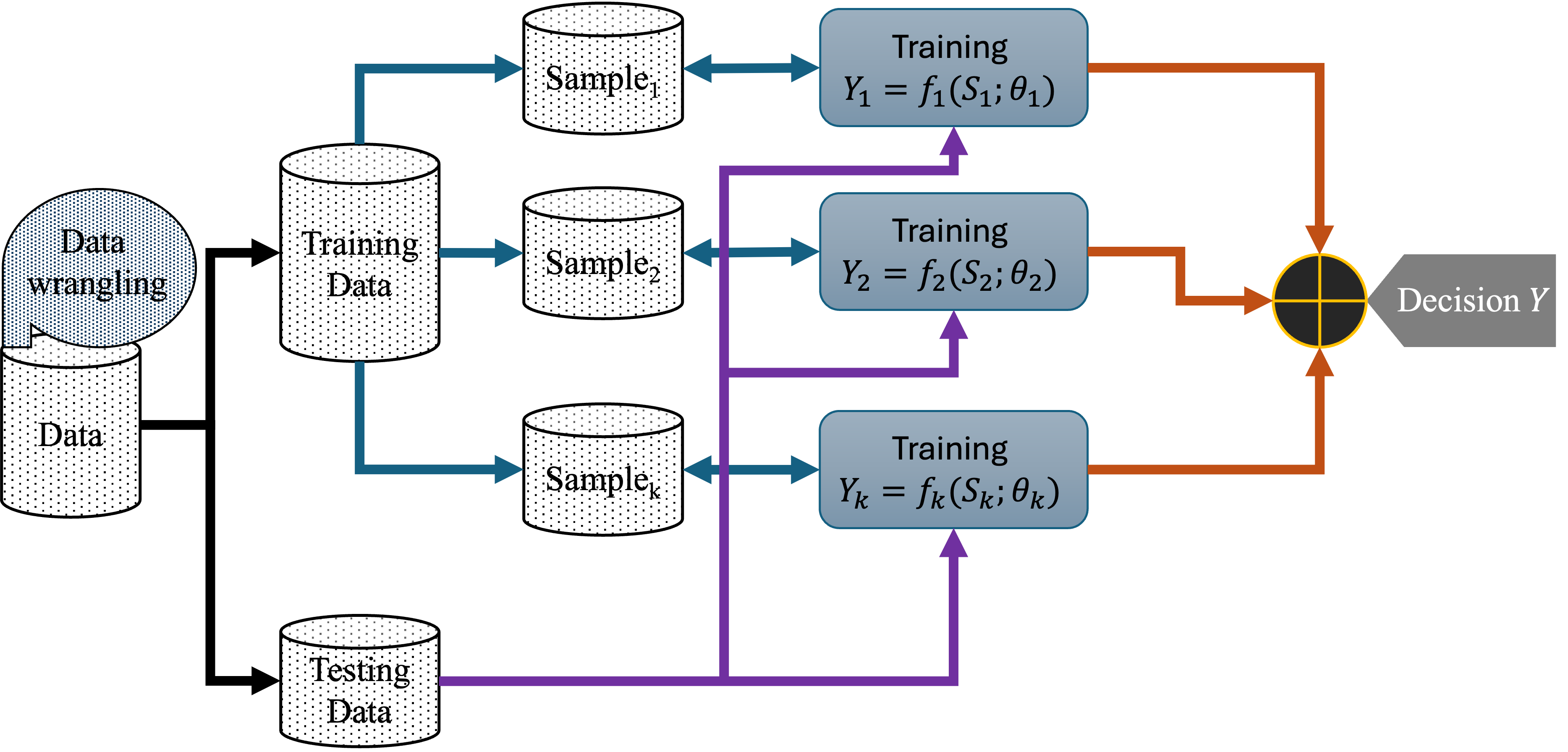}
\caption{The figure shows the necessary components for an ensemble machine learning process. The process starts with data. The feature engineering process extracts important features and prepares data for the next level of operations. The data is split into two disjoint subsets: training and testing data. The training dataset is split into $k$ sample datasets. Based on the ensemble model algorithm, each subsample of data trains a weak learner and finally aggregates the outputs (decisions) to make a combined decision.}
\eefs{ml-details-ensemble}

In many scenarios, combined decisions from multiple models have better prediction probability than a single model. FIG.~\ref{fig:ml-details-ensemble} shows the basic architecture of an ensemble model. Each model in the ensemble model is a weak learner, and the combined decision of all weak learners produces a strong decision. A weak learner is a machine learning model with better prediction power than random choice. Ensemble models are categorized into three parts: a) Parallel, b) Sequential, and c) Composite. Bagging and Voting ensemble models are parallel models where many weak learners are trained parallelly on different data samples. The combined decision is used as the model's decision. Random forest is a popular bagging algorithm. Boosting algorithms like ``Gradient boosting" or ``XGBoost" also use many weak learners, but they train the model sequentially.

\subsection{Tree-based ensemble methods: random forest, gradient boosted decision tree, and extreme gradient boost}\label{sec:MLMethods}

The random forest (RF) regressor is the simplest ensemble method, in which each regressor is a decision tree trained on a random sample $S_i$ of the data, and each tree has a unit weight in the ensemble~\cite{breiman19991}. Therefore, random forests are computationally efficient and have low variance.
The number of trees and the depth of the tree are two hyperparameters of an RF regressor. By selecting a random subset of features for each tree and training each tree on a bootstrapped dataset sample, this randomness helps prevent overfitting and decorrelates the individual trees, resulting in a more accurate and stable predictor.

As proposed in~\cite{friedman2001greedy}, the gradient boosted decision tree (GBT)~\cite{boehmke2019gradient} is another tree-based ensemble method where several decision trees are used as weak learners. Together, all the weak learners form a strong and effective regressor. Initially, each sample is weighted equally, and then the first weak regressor trains the first sample, and so on. After learning, one reduces the weight of the correctly predicted samples and increases the weight of the mistaken samples. Residuals are computed from the mistaken samples, and a weak regressor is trained based on the previous weak regressor's residual error. In this way, GBT can reach the prediction target by decreasing the residual error in the training process.

Extreme gradient boost (XGB) regressor~\cite{chen2016xgboost} is an improved version of the GBT regressor regarding computing speed, generalization, and scalability. Its efficient implementation uses parallel and distributed computing, allowing faster training times than traditional gradient-boosting algorithms. One of XGB's critical features is its ability to handle missing data and outliers effectively. XGB enhances model generalization, mitigates overfitting, and provides robustness in noisy or incomplete datasets by employing a sparsity-aware algorithm and incorporating regularization terms.

In the following, we discuss the application of these ML algorithms on the lattice data, which we call the ``experimental data" or ``experimental results" for the ML analyses.

\subsection{Experimental results} 
\subsection*{Matrix elements for the polarized gluon correlation function}
LQCD dataset associated with the polarized gluon distribution consists of a vector of $1901$ data points ($\Dl \mathfrak{M}$) for each pair of momentum $p_z$ and $\om$. 1901 is the number of configurations used in the LQCD calculations~\cite{HadStruc:2022yaw,Khan:2022vot}. By each pair of momentum, we refer to the relation between two-$p_z$ datasets in Eq.~\eqref{eq:master}. We computed an aggregate dataset $\mathcal{D}_r$ such that $\mathcal{D}_r=\{(p,w,\mu_r(w),\sigma_r(w)): \mu_r(w)\leftarrow {\rm mean}(\Dl \mathfrak{M}(w)), \sigma_r(w)\leftarrow {\rm std.deviation}(\Dl \mathfrak{M}(w)), p\in p_z, w\in \om\}$. Fig.~\ref{fig:polarized-data} shows the $\Dl \mathfrak{M}(\om)$ vs $\om$,  and the points are simply connected for each $p_z$. $\Dl \mathfrak{M}(\om)=0$ when $\om=0$. When $\om$ increases, different curves (for different $p_z$) have different contributions from the contamination term, as seen from Eq.~\eqref{eq:Ipform1}.  $\Dl \mathfrak{M}$ curves move towards the phenomenological ITD~\cite{Sufian:2020wcv} as $p_z$ increases and show features of decreasing around $\om  \geq 6$ at $p_z=2.46$ GeV within uncertainty. The challenge of removing the contamination term $(m_p^2/p_z^2) \om \Dl{\mathcal{M}}_{pp}$ and obtaining a nonzero signal can be understood from the smallness in the phenomenological gluon helicity ITD, as noted in~\cite{Sufian:2020wcv, Khan:2022vot}. The pattern of the curve outside the available lattice data is unclear. Due to the small number of observations, any ML model will struggle with underfitting and the uncertainty of the extrapolated data will be large.

To overcome this limitation, we incorporated physics constraints and knowledge from the phenomenological ITD from~\cite{Sufian:2020wcv} (or alternatively can be used from~\cite{Ball:2017nwa}) along with the aggregate dataset ($\mathcal{D}_r$) to generate synthetic data. We have used two phenomenological ITDs from different fits in~\cite{Sufian:2020wcv}, as shown in Fig.~\ref{fig:phenoITD}. The magnitude of the ML-generated data is not affected by those of the phenomenological ITDs. The knowledge of the phenomenological ITDs outside the LQCD data domain helps to determine the shape of the ML-generated ITD at large $\om$ after eliminating the contamination term. Also, as discussed in~\cite{Khan:2022vot}, within the current statistics, the matching effects on $\Dl \mathfrak{M}_g(\om)$ in the $\ms$-scheme is marginal within uncertainty. One can convert the phenomenological ITD data to the reduced pseudo-ITD data using the matching formula in~\cite{Balitsky:2021cwr} and verify this. Therefore, we assume the shape of the phenomenological ITD is similar to that of the $\Dl \mathfrak{M}_g(\om)$ where LQCD data is unavailable. One can numerically check that synthetic data using either the fit-1 or fit-2 curves in Fig.~\ref{fig:phenoITD} results in the same curves from the lattice data. For example, as we will see, the two phenomenological curves have maximum values, $\Del \mathcal{I}_g \approx 0.4$ and $0.2$. In contrast, the contamination-free $\Dl \mathfrak{M}_g$ will have a maximum value of around $0.12$ by fitting the LQCD data along with the synthetic data.  

\befs 
\centering
\includegraphics[scale=0.72]{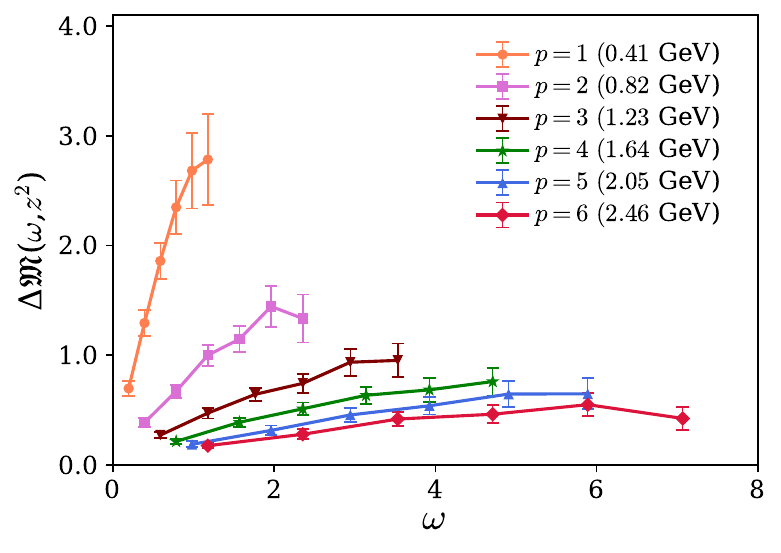}
\caption{\label{fig:lattdaat}
 $\Dl \mathfrak{M}$ values  from the LQCD calculation against each momenta $p_z$ and $\om$. The mean values are connected with lines. $p_z$ values are denoted by simply $p=n$, where $n=1,2,\cdots 6$ according to discrete lattice momenta $p_n = \frac{2\pi n}{La}$. The corresponding values in the physical unit of GeV are shown in parentheses.}     
\eefs{polarized-data}

\subsubsection*{Generating synthetic data}
Let $\Dl \mathfrak{M}_i(\om)$ be the data points associated with $p_i$ curve (from Fig.~\ref{fig:polarized-data}). Utilizing Eq.~\eqref{eq:master}, for each $w\in\om$,  $\Dl \mathfrak{M}_g(w)$ can be obtained as follows:
\bea\label{eq:IvsM}
     \Dl \mathfrak{M}_g = \frac{j^2\Dl \mathfrak{M}_j-i^2 \Dl \mathfrak{M}_i}{j^2-i^2}&=\frac{k^2 \Dl \mathfrak{M}_k-i^2 \Dl \mathfrak{M}_i}{k^2-i^2}, (j\neq k)>i>0 .
 \eea
Let,
\begin{align*}
    \mathcal{A} &= \frac{j^2}{j^2-i^2}, \quad
    1-\mathcal{A} = \frac{-i^2}{j^2-i^2}\\
    \mathcal{B} &= \frac{k^2}{k^2-i^2}, \quad
    1-\mathcal{B} = \frac{-i^2}{k^2-i^2}\\
    \mathcal{C} &= \frac{i^2}{k^2}.
\end{align*}
After replacing the expressions in Eq.~\eqref{eq:IvsM} we get:
\begin{align}\label{eq:twoM}
    \Dl \mathfrak{M}_g = \mathcal{A} \Dl \mathfrak{M}_j+(1-\mathcal{A})\Dl \mathfrak{M}_i &= \mathcal{B} \Dl\mathfrak{M}_k+(1-\mathcal{B}) \Dl\mathfrak{M}_i,\nonumber\\
    \Dl\mathfrak{M}_k &= \frac{\mathcal{A} \Dl\mathfrak{M}_j+(\mathcal{B}-\mathcal{A}) \Dl\mathfrak{M}_i}{\mathcal{B}}\nonumber\\
    &= \mathcal{C} \Dl\mathfrak{M}_i + (1-\mathcal{C})\Dl \mathfrak{M}_g .
\end{align}
From Eq.~\eqref{eq:twoM}, for a specific $\Dl \mathfrak{M}_g(\om)$, values at two known momenta are sufficient to derive the values of $\Dl \mathfrak{M}_g(\om)$ across all different momenta. The equation also states that $\Dl \mathfrak{M}_g(\om)$ is a weighted average of two $\Dl\mathfrak{M}(\om)$, and the shape of $\Dl \mathfrak{M}_g(\om)$ and $\Dl\mathfrak{M}(\om)$ are correlated. Once we know the pattern of $\Dl \mathfrak{M}_g(\om)$, we can derive the pattern of momentum-dependent matrix elements. 

\befs 
\centering
\includegraphics[scale=0.72]{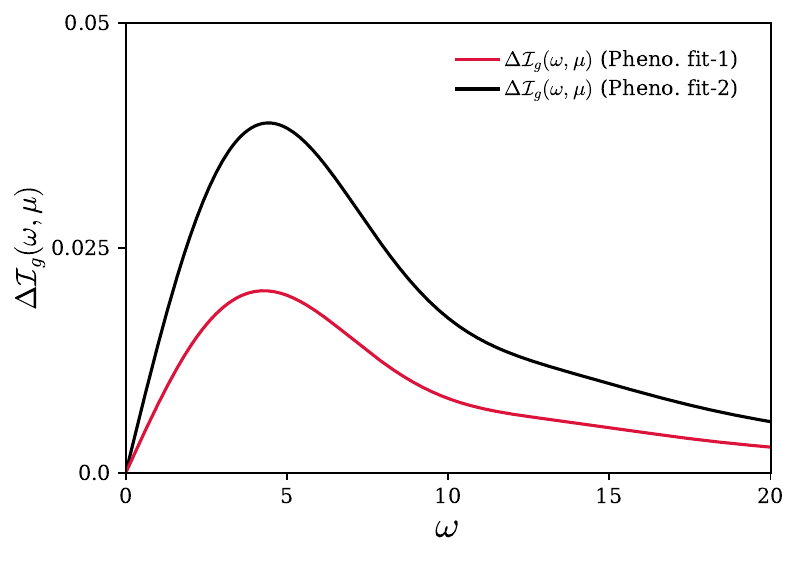}
\caption{Phenomenological shapes of $\Dl \mathcal{I}_g(\om, \mu)$ with respect to $\om$ from~\cite{Sufian:2020wcv} in the $\ms$ scheme at $\mu=2$ GeV. As discussed in the text, we focus here on the central lines of two different fit results from~\cite{Sufian:2020wcv} to determine the shape of the ITD. }
\eefs{phenoITD}

In the following, we will present a specific case of ML, the gradient descent algorithm~\cite{ruder2016overview}. Fig.~\ref{fig:phenoITD} shows the pattern of the phenomenological $\Dl \mathcal{I}_g(\om)$, which is used to derive $\Dl \mathfrak{M}(\om)$ using the
gradient descent algorithm. Gradient descent is an optimization algorithm that finds the minimum or maximum in the loss function. A loss function must be either convex or concave to have a minimum or maximum value, respectively. In our solution, we have $\Dl \mathfrak{M}_g(\om)$ from Eq.~\eqref{eq:twoM}, and $\Dl \mathcal{I}_g(\om)$ from Fig.~\ref{fig:phenoITD}. The convex loss function is defined as $\mathcal{L}\equiv (\Dl \mathfrak{M}_g(\om)-\Dl \mathcal{I}_g(\om))^2$. Derivation of the loss function concerning $\Dl
\mathfrak{M}_g(\om)$ denoted by $\pdv{\mathcal{L}}{\Dl \mathfrak{M}_g(\om)}$ indicates a slope at $\Dl \mathfrak{M}_g(\om)$ on the loss curve. We update the $\Dl \mathfrak{M}_g(\om)$ with respect to the derivative $\Dl \mathfrak{M}_g(\om)=\Dl \mathfrak{M}_g(\om)-\eta\pdv{\mathcal{L}}{\Dl \mathfrak{M}_g(\om)}$. This ensures the shift of $\Dl \mathfrak{M}_g(\om)$ towards the minimum point. The pace or step size of the update depends on $\eta$, which is called the learning rate. Algorithm~\ref{algo: find-momentum} shows the details
to estimate $\Dl \mathfrak{M}_g(\om)$ by generating the synthetic data $\mathcal{D}_g$. 

\noindent
\begin{minipage}{0.55\textwidth}
\vspace*{-0.13cm}
\begin{algorithm}[H]
    \caption{Find $\Dl \mathfrak{M}(\om)$ values using gradient descent}
    \label{algo: find-momentum}
    \SetKwInOut{Input}{Input}
    \SetKwInOut{Output}{Output}
    \Input{Data $\Dl \mathcal{I}_g (\om)$, index $i$ for $p_i$, index $j$ for $p_j$ }
    \Output{Data $\mathcal{D}_g$}
    $\alpha\leftarrow\frac{j^2}{j^2-i^2}$\\
    $\beta\leftarrow\frac{-i^2}{j^2-i^2}$\\
    Initialize learning rate, $\eta\leftarrow0.01$\\
    $\mathcal{D}_g\leftarrow\emptyset$\\
    \For{each point $(\Dl \mathcal{I}_g,\om)\in \Dl \mathcal{I}_g(\om)$}{
        Initialize, $\Dl \mathfrak{M}_j\leftarrow 0$\\
        Initialize, $\Dl \mathfrak{M}_i\leftarrow n*\Dl \mathcal{I}_g,\quad n>0$\\
        \For{epoch $\in\{1,\dots,100\}$}{
            $\Dl\mathfrak{M}_g\leftarrow\alpha \Dl \mathfrak{M}_j + \beta \Dl \mathfrak{M}_i$\\
            $\mathcal{L}\leftarrow (\Dl \mathcal{I}_g-\Dl\mathfrak{M}_g)^2 = (\Dl \mathcal{I}_g - (\alpha \Dl \mathfrak{M}_j + \beta \Dl \mathfrak{M}_i))^2$ \\
            $\frac{\partial\mathcal{L}}{\partial \Dl\mathfrak{M}_j}\leftarrow -2\times\alpha\times(\Dl \mathfrak{M}_g-\Dl\mathcal{I})$\\
            $\frac{\partial\mathcal{L}}{\partial \Dl\mathfrak{M}_i}\leftarrow -2\times\beta\times(\Dl \mathfrak{M}_g- \Dl\mathcal{I}')$\\
            $\Dl\mathfrak{M}_j\leftarrow \Dl\mathfrak{M}_j-\eta\times\frac{\partial\mathcal{L}}{\partial \Dl\mathfrak{M}_j}$\\
            $\Dl\mathfrak{M}_i\leftarrow \Dl\mathfrak{M}_i-\eta\times\frac{\partial\mathcal{L}}{\partial \Dl\mathfrak{M}_i}$\\
        }
        $\mathcal{G}\leftarrow[\om,\Dl\mathfrak{M}_i,\Dl\mathfrak{M}_j]$\\
        \For{$k\in\{\ell:\ell>i,\ell\neq j\}$}{
            $\gamma\leftarrow\frac{i^2}{k^2}$\\
            $\Dl\mathfrak{M}_k\leftarrow(1-\gamma)\times \Dl \mathcal{I}_g+\gamma\times \Dl\mathfrak{M}_i$\\
            $\mathcal{G}\leftarrow Append(\mathcal{G}, \Dl \mathfrak{M}_k)$\\
        }
        $\mathcal{D}_g\leftarrow Append(\mathcal{D}_g, \mathcal{G})$\\
    }
\end{algorithm}
\end{minipage}
\hspace{1ex}

\befs 
\centering
\includegraphics[scale=0.3]{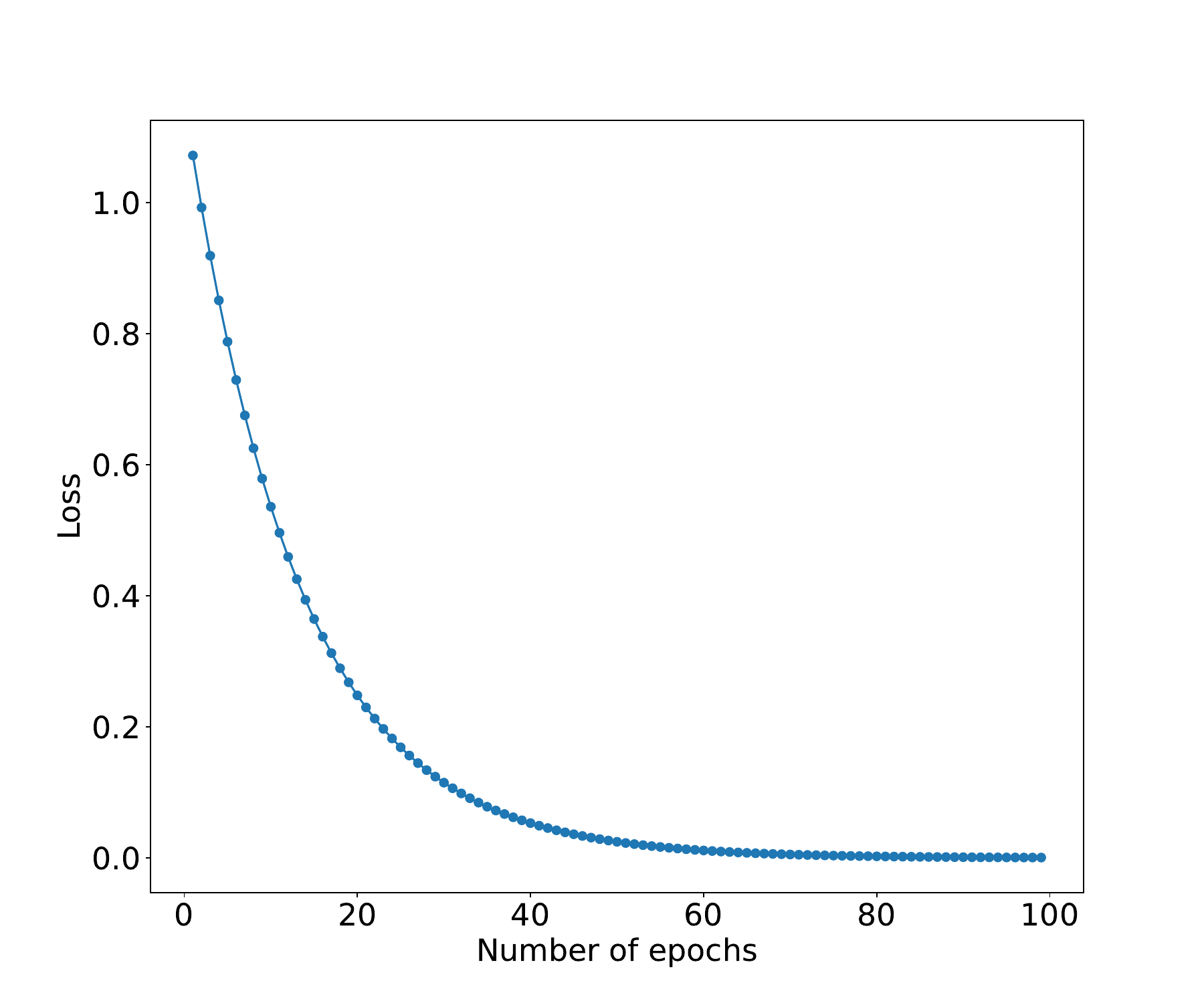}
\caption{\label{fig:lossfunc} 
 The loss across the epochs.}  
 \label{fig:loss} 
 \eefs{mockdemocn}

\noindent\paragraph{\bf{Algorimic details:}} The details of the algorithm are as follows:
\begin{itemize}[leftmargin=6em]
    \item [Step 1:] Our algorithm takes the values of the curve plotted in Fig.~\ref{fig:polarized-data} at two $p$ values ($i$ for $p_i$ and $j$ for $p_j$ are randomly selected). 
    
    \item [Step 2:] We created two variables based on the index of $p$ at line $1$ and $2$ as shown in Algorithm~\ref{algo: find-momentum}. These are the weights to compute $\Dl \mathfrak{M}_g$. 
    
    \item [Step 3:] We initialize the learning rate at $0.01$. The value of the learning rate is critical to reach the minimum point. If the value is very small, it will take a long time to reach the minimum, and if the value is large, it will jump from one point to another and miss the minimum point. Conventionally, the learning rate ranges from $e^{-7}$ to $e^{-1}$. 
    
    \item [Step 4:] The outer loop in line $5$ runs for each data point of Fig.~\ref{fig:phenoITD}. At each iteration of this loop, we initialize the $\Dl\mathfrak{M}$ and pass that to another loop (in line $8$) where, at each iteration, the $\Dl\mathfrak{M}$ are adjusted based on the gradient of the $\Dl\mathfrak{M}$. At first, we initialize $\Dl \mathfrak{M}_g$ as the linear combination of $\Dl\mathfrak{M}_i$ and $\Dl\mathfrak{M}_j$. The derivative of the loss function with respect to $\Dl\mathfrak{M}_i$ is used to update the value of $\Dl\mathfrak{M}_i$ for the next iteration. Similarly, the derivative of the loss function with respect to $\Dl\mathfrak{M}_j$ is used to update the value of $\Dl\mathfrak{M}_j$ for the next iteration. This update accelerates the value of the variables ($\Dl\mathfrak{M}_i$ and $\Dl\mathfrak{M}_j$) towards its minimum and reduces the loss close to zero (shown in Fig.~\ref{fig:loss}).
    
    \item [Step 5:] We adjusted the initialization of $\Dl\mathfrak{M}_i$ and $\Dl\mathfrak{M}_j$ to achieve the best overlapping between the real and the estimated values (shown in Fig.~\ref{fig:fitted-M(w)}). To optimize the overlap and adhere to the relation between two $p_z$ datasets, it becomes apparent that only $z\leq 4a (= 0.36~{\rm fm})$  LQCD matrix elements can be incorporated into the fit while satisfying Eq.~\eqref{eq:twoM}.~\footnote{This observation aligns with  the findings in~\cite{Khan:2022vot}, indicating that LQCD matrix elements at large physical separations may involve substantial power corrections, rendering them unsuitable for determining the light-cone correlation function associated with the gluon helicity PDF. Further discussion on this topic follows in the subsequent section of the manuscript.} 
    
    \item [Step 6:] The latter part of the algorithm is used to generate the other values {$\Dl\mathfrak{M}_k$} based on the $i$-th and $j$-th  values $\Dl\mathfrak{M}_i$ and $\Dl\mathfrak{M}_j$ associated with the $i$-th and $j$-th momenta. We also created a synthetic dataset where for each $p\in p_z$, we have $200~\Dl\mathfrak{M}(\om)$ values. 
\end{itemize}

\befs
\centering
    \includegraphics[width=0.8\textwidth]{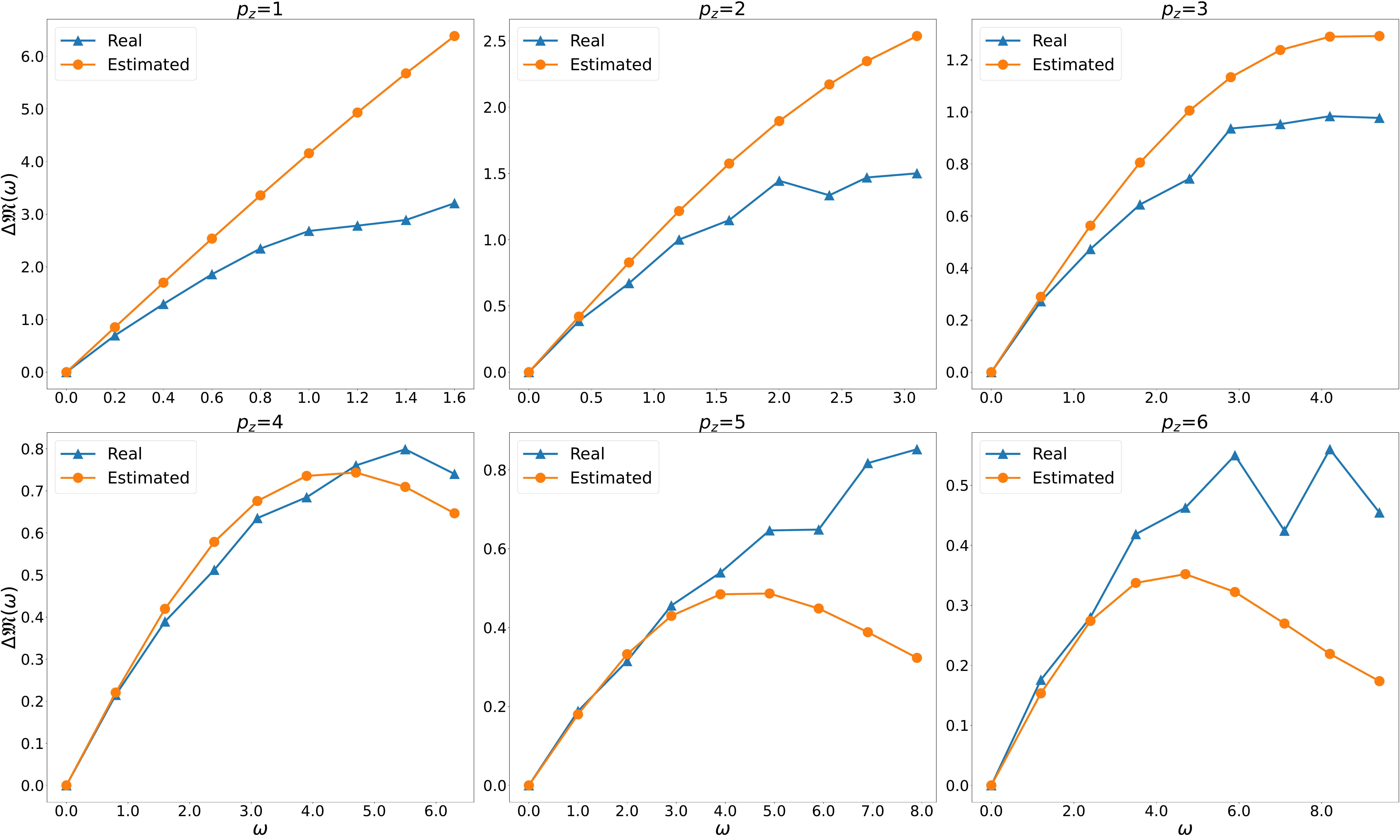}
    \caption{Estimated $\Delta \mathfrak{M}(\om)$ for all momenta using algorithm~\ref{algo: find-momentum}.  At this stage, only mean values have been estimated. The error bars will be determined after estimating the standard deviation and randomly picking values from a normal distribution. Note that $z\geq 5a$ data points are not used in the fit. $p_z$ values are denoted by simply $p=n$, where $n=1,2,\cdots 6$ according to discrete lattice momenta $p_n = \frac{2\pi n}{La}$. By the blue-colored ``real" data, we refer to the matrix elements obtained from the LQCD calculation, while the orange-colored ``estimated" data refers to the ML-estimated synthetic data.}
\eefs{fitted-M(w)}

\befs
\centering
    \includegraphics[width=0.8\textwidth]{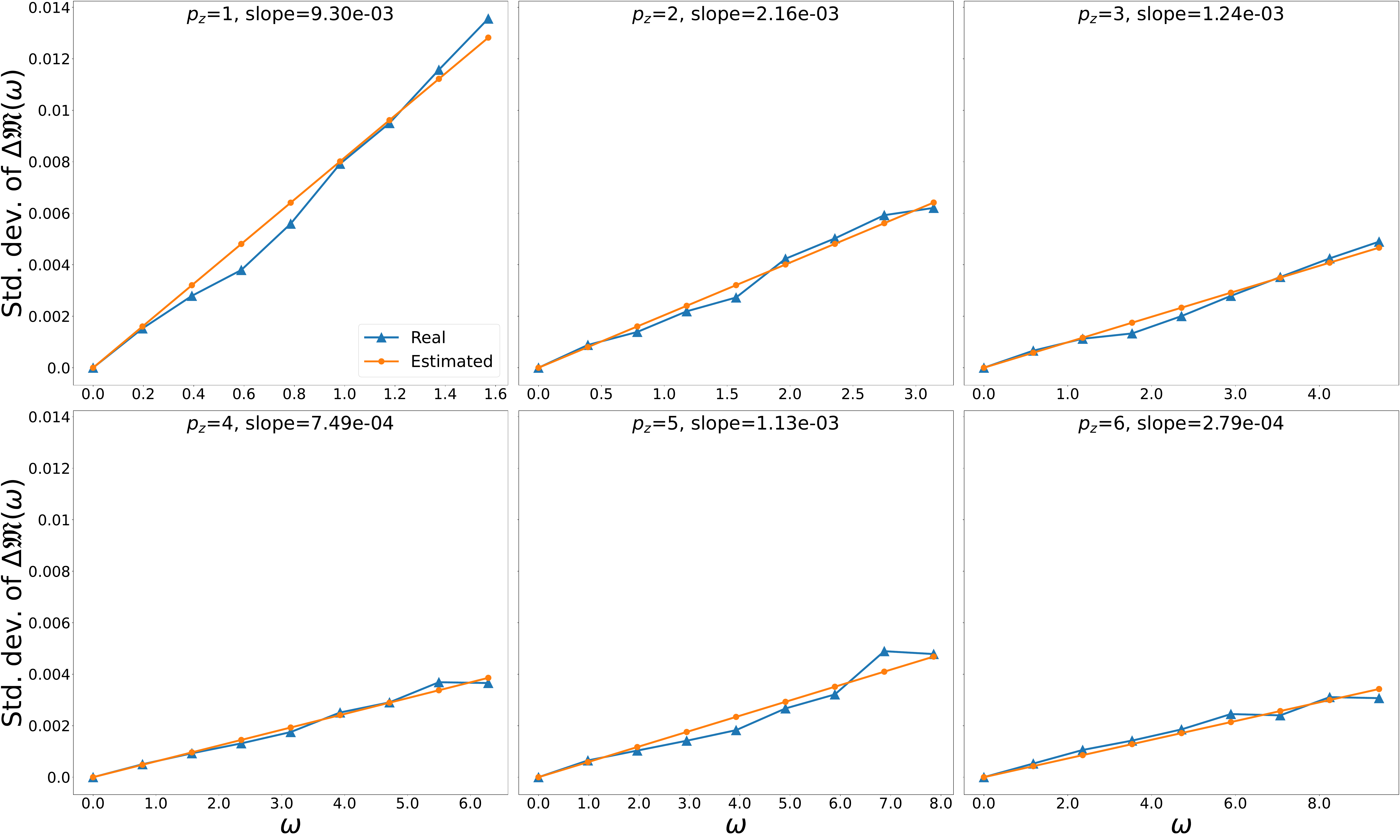}
    \caption{The standard deviation and the best-fitted linear regression model on the standard deviation data are shown here for each $p$. $p_z$ values are denoted by simply $p=n$, where $n=1,2,\cdots 6$ according to discrete lattice momenta $p_n = \frac{2\pi n}{La}$. } 
\eefs{std-lr}

\noindent\paragraph{\bf{Standard deviation prediction:}} For each ($p\in p_z, w\in\om$), we have $1901$ values of $\Dl\mathfrak{M}(\om)$ in the dataset $\mathcal{D}_r$, from which we can compute mean and standard deviation of $\Dl\mathfrak{M}(\om)$. The blue line in Fig.~\ref{fig:std-lr} shows the pattern of the standard deviation. The pattern shows linear behavior, and we estimated the best-fitted linear regression model ($\mathfrak{f}_{LR}(p,w)$). The orange line in Fig.~\ref{fig:std-lr} shows the best-fitted linear regression model. For each ($x\in p_z, y\in\om$) in the generated dataset $\mathcal{D}_g$ (using Algorithm~\ref{algo: find-momentum}), we used the best fitted linear regression ($\mathfrak{f}_{LR}(x,y)$) to estimate the standard deviation. The estimated standard deviation is used to generate the $\Dl\mathfrak{M}$ for the dataset $\mathcal{D}_g$.

\noindent\paragraph{\bf{Data generation:}} From the real dataset $\mathcal{D}_r$, we intend to understand the distribution of $\Dl\mathfrak{M}(\om)$ for each ($p\in p_z, w\in\om$). Fig.~\ref{fig:dist_m} shows the mean and median plot, and having all the points on the same diagonal line indicates that the distribution of $\Dl\mathfrak{M}(\om)$ is normal for each ($p\in p_z, w\in\om$). This knowledge is  utilized to estimate $\Dl\mathfrak{M}(\om)$ for dataset $\mathcal{D}_g$ as follows:

\begin{enumerate}
    \item For each ($p\in p_z, w\in\om$), the estimated mean ($\mu_{g,\om}$) and standard deviation ($\sigma_{g,\om}$) are used to generate a normal distribution $\mathcal{N}(\mu_{g,\om}, \sigma_{g,\om})$.
    \item We randomly generate $1901$ values from the distribution like $\{x_1, x_2, \dots, x_{1901}\}$ where $x_i\sim\mathcal{N}(\mu_{g,\om}, \sigma_{g,\om})$.
\end{enumerate}

\befs
\centering
    \includegraphics[width=0.4\textwidth]{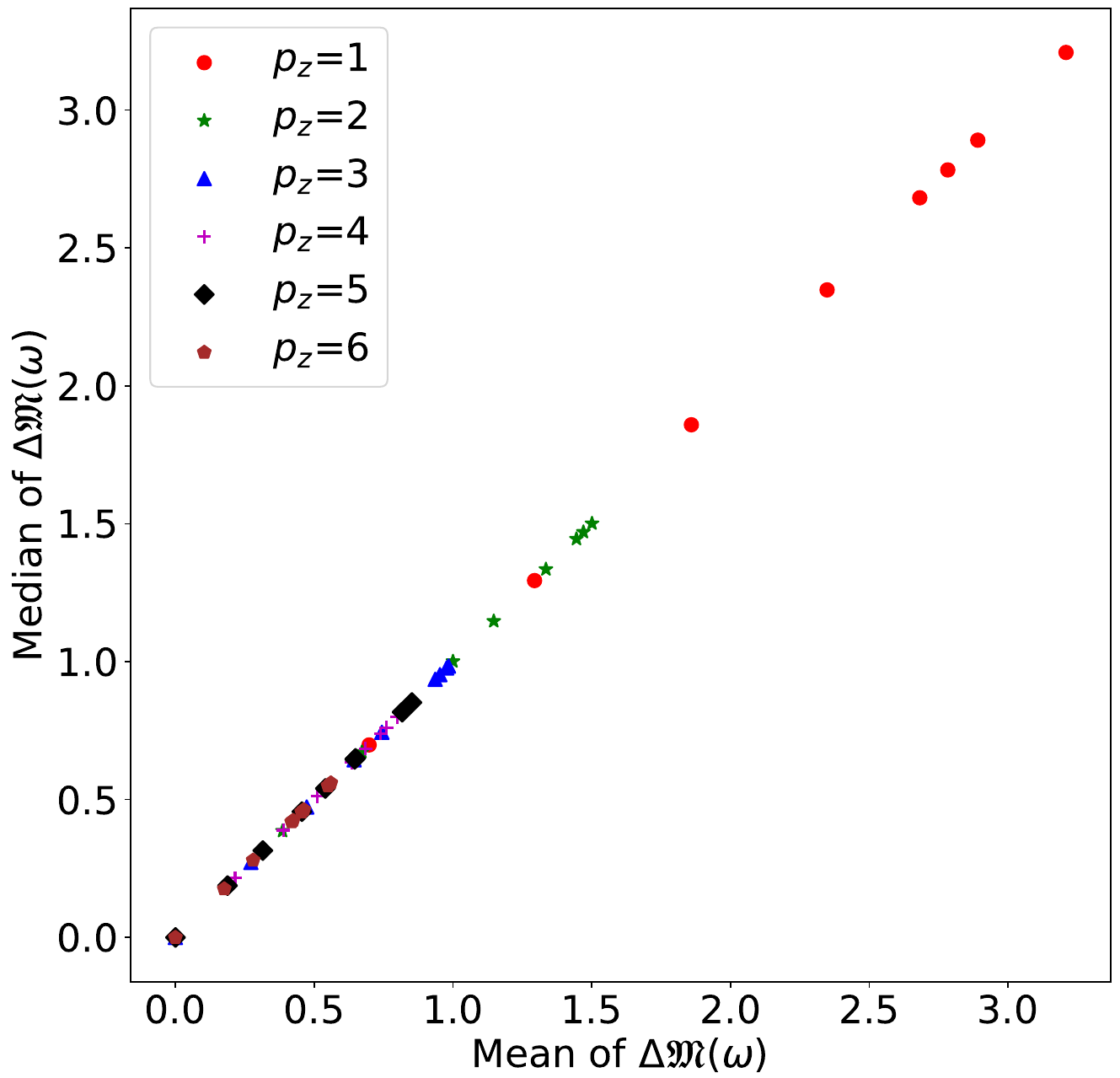}
    \caption{For each ($p, \om$), the mean and median values of $\Delta \mathfrak{M}(\om)$ are plotted here. All the points are on the diagonal line, indicating that the mean and median values are the same. The $\Delta \mathfrak{M}(\om)$ follows a normal distribution. }. 
\eefs{dist_m}

\noindent\paragraph{\bf{Machine learning model evaluation:}} 
The generated dataset $\mathcal{D}_g$ contains six $p$-values, $200$  values of $\om$ for each $p$, and $1901$  values of $\Dl\mathfrak{M}(\om)$ for each ($p,\om$). Alternatively, we can say that the dataset contains six $p$-values, and for each $p$, there are $1901$ sequences of ($\om, \Dl\mathfrak{M}(\om)$) where the length of each sequence is $200$. As $\om$ increases, the non-linear nature of $\Dl\mathfrak{M}(\om)$ indicates the influence of other hidden variables (say it is $\mathcal{a}$), formulated as $\Dl\mathfrak{M}(\om)=\mathfrak{f}(\om, \mathcal{a})$. In our experiment, we used the prior knowledge of ($\om_{t-1}, \Dl\mathfrak{M}(\om_{t-1})$) as the hidden variable, which reduces the problem to a known problem called Bayesian probability to predict $\Dl\mathfrak{M}(\om)$ using prior observations. We formulated the prediction problem as follows:
\begin{align}\label{eq:MLPred}
    \mathbb{P}_r(\Dl\mathfrak{M}_\ell|\om_\ell,\Dl\mathfrak{M}_{\ell-1},\om_{\ell-1},\dots,\Dl\mathfrak{M}_{\ell-k},\om_{\ell-k}),
\end{align}
where $\ell\geq k$ and $\Dl\mathfrak{M}_{\ell}\equiv\Dl\mathfrak{M}(\om_\ell)$. For our experiment, we set $k=5$. We prepared training and testing datasets to apply our prediction problem. 

\begin{table}[ht!]
    \centering
    \begin{tabular}{lllllllllll|l}\hline
        $h_1$ & $h_2$ & $h_3$ & $h_4$ & $h_5$ & $h_6$ & $h_7$ & $h_8$ & $h_9$ & $h_{10}$ & $h_{11}$ & Target\\\hline
        $\om_1^1$ & $\Dl\mathfrak{M}_1^1$ & $\om_1^2$ & $\Dl\mathfrak{M}_1^2$ & $\om_1^3$ & $\Dl\mathfrak{M}_1^3$ & $\om_1^4$ & $\Dl\mathfrak{M}_1^4$ & $\om_1^5$ & $\Dl\mathfrak{M}_1^5$ & $\om_1^6$ & $\Dl\mathfrak{M}_1^6$\\
        $\om_1^2$ & $\Dl\mathfrak{M}_1^2$ & $\om_1^3$ & $\Dl\mathfrak{M}_1^3$ & $\om_1^4$ & $\Dl\mathfrak{M}_1^4$ & $\om_1^5$ & $\Dl\mathfrak{M}_1^5$ & $\om_1^6$ & $\Dl\mathfrak{M}_1^6$ & $\om_1^7$ & $\Dl\mathfrak{M}_1^7$\\
        \multicolumn{11}{c}{$\dots$} & \\
        $\om_1^{195}$ & $\Dl\mathfrak{M}_1^{195}$ & $\om_1^{196}$ & $\Dl\mathfrak{M}_1^{196}$ & $\om_1^{197}$ & $\Dl\mathfrak{M}_1^{197}$ & $\om_1^{198}$ & $\Dl\mathfrak{M}_1^{198}$ & $\om_1^{199}$ & $\Dl\mathfrak{M}_1^{199}$ & $\om_1^{200}$ & $\Dl\mathfrak{M}_1^{200}$\\
        $\om_2^1$ & $\Dl\mathfrak{M}_2^1$ & $\om_2^2$ & $\Dl\mathfrak{M}_2^2$ & $\om_2^3$ & $\Dl\mathfrak{M}_2^3$ & $\om_2^4$ & $\Dl\mathfrak{M}_2^4$ & $\om_2^5$ & $\Dl\mathfrak{M}_2^5$ & $\om_2^6$ & $\Dl\mathfrak{M}_2^6$\\
        \multicolumn{11}{c}{$\dots$} & \\
        $\om_{1901}^{195}$ & $\Dl\mathfrak{M}_{1901}^{195}$ & $\om_{1901}^{196}$ & $\Dl\mathfrak{M}_{1901}^{196}$ & $\om_{1901}^{197}$ & $\Dl\mathfrak{M}_{1901}^{197}$ & $\om_{1901}^{198}$ & $\Dl\mathfrak{M}_{1901}^{198}$ & $\om_{1901}^{199}$ & $\Dl\mathfrak{M}_{1901}^{199}$ & $\om_{1901}^{200}$ & $\Dl\mathfrak{M}_{1901}^{200}$\\
        \hline
    \end{tabular}
    \caption{The structure of the training data}
    \label{tab:train_data_polarized}
\end{table}

For each $p\in p_z$, we have a sequence of $(\om_i^j, \Dl\mathfrak{M}_i^j)$ for $1\leq i\leq1901, 1\leq j\leq200$ and TABLE~\ref{tab:train_data_polarized} shows the formation of training data using this sequence of data. The training data is passed through the ML model where $10$-fold cross-validation is used to find the best model. During this cross-validation step, the overall dataset splits into $10$ non-overlapping datasets. At each time, one portion is reserved for testing and the rest $9$ fragments of the data are merged for training a model. This process runs $10$ times to test all the $10$ fragments of the data. The best-fitted model is used to generate new $\Dl\mathfrak{M}$ data. We applied three ensemble models named gradient-boosted tree (GBT), random forest (RF), and extreme gradient-boosted tree (XGB) (described in Section~\ref{sec:MLMethods}). These models' root mean square error (RMSE) in training data  are $0.0427$ for the GBT, $0.0404$ for the RF, and $0.0410$ for the XGB. Based on the training error, the RF is the best-performing model.

\noindent\paragraph{\bf{Estimating $\Dl\mathfrak{M}(\om)$ using the best model:}} For each ($p(i)\in p_z, w(i)\in\om$) where $1\leq i\leq1901$, we used sliding--window process to estimate $\Dl\mathfrak{M}(\om)$ as follows:
\begin{enumerate}
    \item We initialize the window with the first five values of $\om\in\{0.1,0.2,0.3,0.4,0.5\}$ and corresponding $\Dl\mathfrak{M}(\om)$ from  the dataset $\mathcal{D}_r$. For any missing $\Dl\mathfrak{M}(\om)$, we use the corresponding $\Dl\mathfrak{M}(\om)$ from  the dataset $\mathcal{D}_g$. 
    \item Using the prediction model  Eq.~\eqref{eq:MLPred}, we compute $\Dl\mathfrak{M}(\om)$ for $\om=0.6$. 
    \item We remove the leftmost entry $(\om, \, \Dl\mathfrak{M}(\om))$ of the window and insert the predicted ($\om=0.7, \, \Dl\mathfrak{M}(\om=0.6))$ to the window to compute the next $\Dl\mathfrak{M}(\om=0.7)$. We repeat the same process to estimate $\Dl\mathfrak{M}(\om)$ for larger values of $\om$.
\end{enumerate}
Fig.~\ref{fig:generated_I} (A--C) shows the estimated $\Dl\mathfrak{M}(\om)$ for larger values of $\om$. The curves for each $p$ are smoother using RF as the predictor.

\noindent\paragraph{\bf{Reconstruction of $\Dl\mathfrak{M}_g(\om)$}:} Fig.~\ref{fig:generated_I} (D--F) represents the reconstruction of $\Dl\mathfrak{M}_g(\om)$ using the predicted $\Dl\mathfrak{M}(\om)$. Each curve corresponds to an ML model. The RF model predicts $\Dl\mathfrak{M}(\om)$ more precisely compared to the other two models, which is demonstrated by the smoothness of the $\Dl\mathfrak{M}_g(\om)$ curve.

\befs
\centering
    \includegraphics[width=\textwidth]{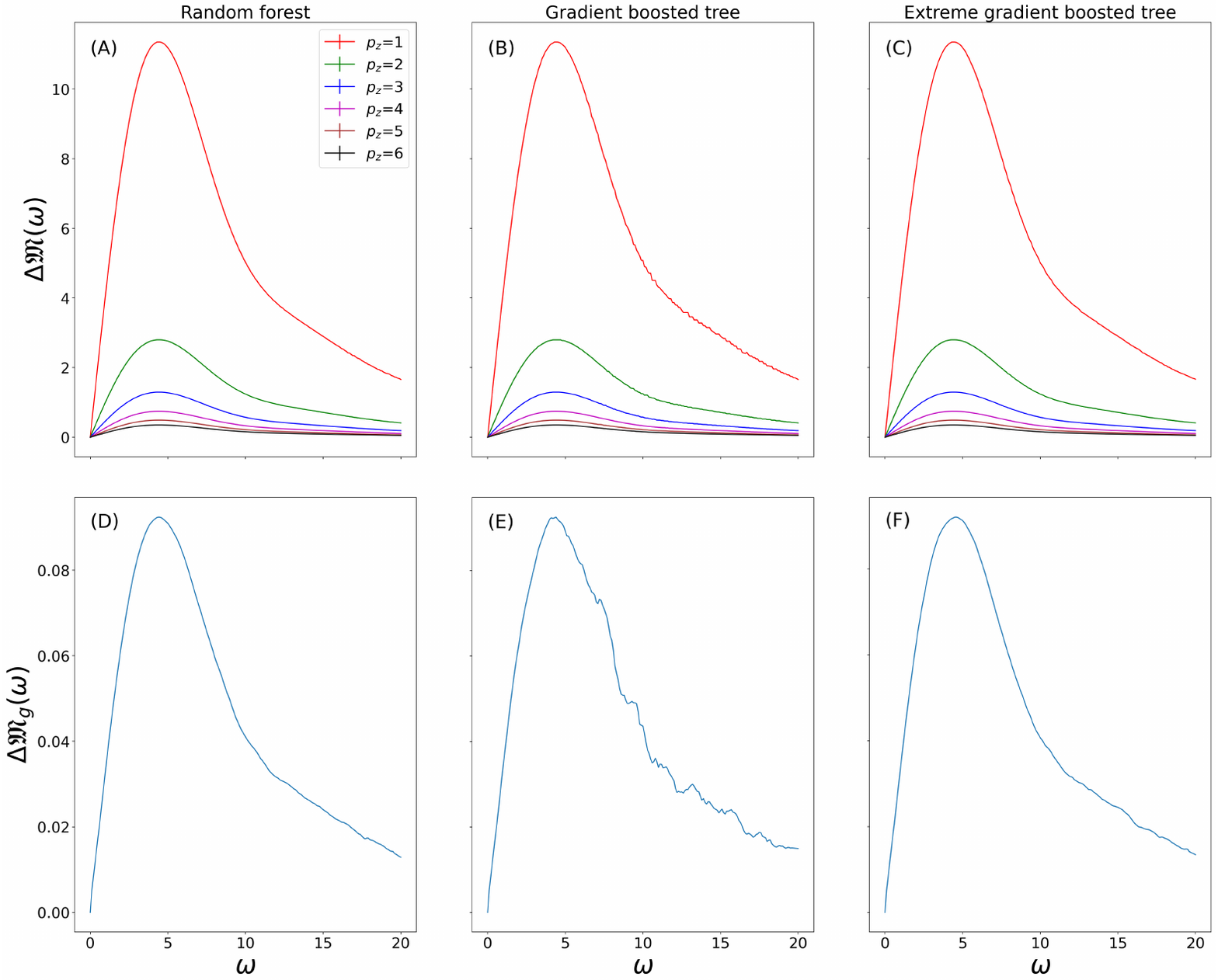}
    \caption{The reconstruction of $\Dl\mathfrak{M}_g(\om)$ from the  $\Dl\mathfrak{M}(\om)$ generated using three machine learning models. The $\Dl\mathfrak{M}(\om)$ generated using the Random forest model produces smoother $\Dl\mathfrak{M}_g(\om)$. Based on the training error, the RF was determined to be the best-performing model.  Here, only the mean values have been shown.} 
\eefs{generated_I}

\subsection*{Matrix elements for the unpolarized gluon correlation function}

We now consider the matrix elements associated with the reduced pseudo-ITD distribution $\mathfrak{M}$~\cite{Balitsky:2019krf,HadStruc:2021wmh}. These matrix elements associated with the unpolarized gluon distribution consist of a vector of $351$  $\mathfrak{M}$ data for each pair of momentum $p_z$ and $\om$, which is used to compute an aggregate dataset $\mathcal{D}_{r}$ such that $\mathcal{D}_{r}=\{(p,w,\mu_r(w),\sigma_r(w)): \mu_r(w)\leftarrow {\rm mean}(\mathfrak{M}(w)), \sigma_r(w)\leftarrow {\rm std.deviation}(\mathfrak{M}(w)), p\in p_z, w\in \om\}$. The number 351 is the number of configurations used in the LQCD calculation~\cite{HadStruc:2021wmh}. Fig.~\ref{fig:unpolarized-data} shows the $\mathfrak{M}(\om)$ vs $\om$, and the points are connected for each $p_z$. $\mathfrak{M}(\om)=1$ when $\om=0$.~\footnote{As in~\cite{HadStruc:2021wmh}, for a comparison with the light-cone ITD in the $\ms$-scheme, the lattice ITD will be normalized using the gluon momentum fraction $\langle x\rangle_g (\mu= 2 \rm{GeV})=0.427$ from~\cite{Alexandrou:2020sml}. }

\befs
\centering
    \includegraphics[width=0.4\textwidth]{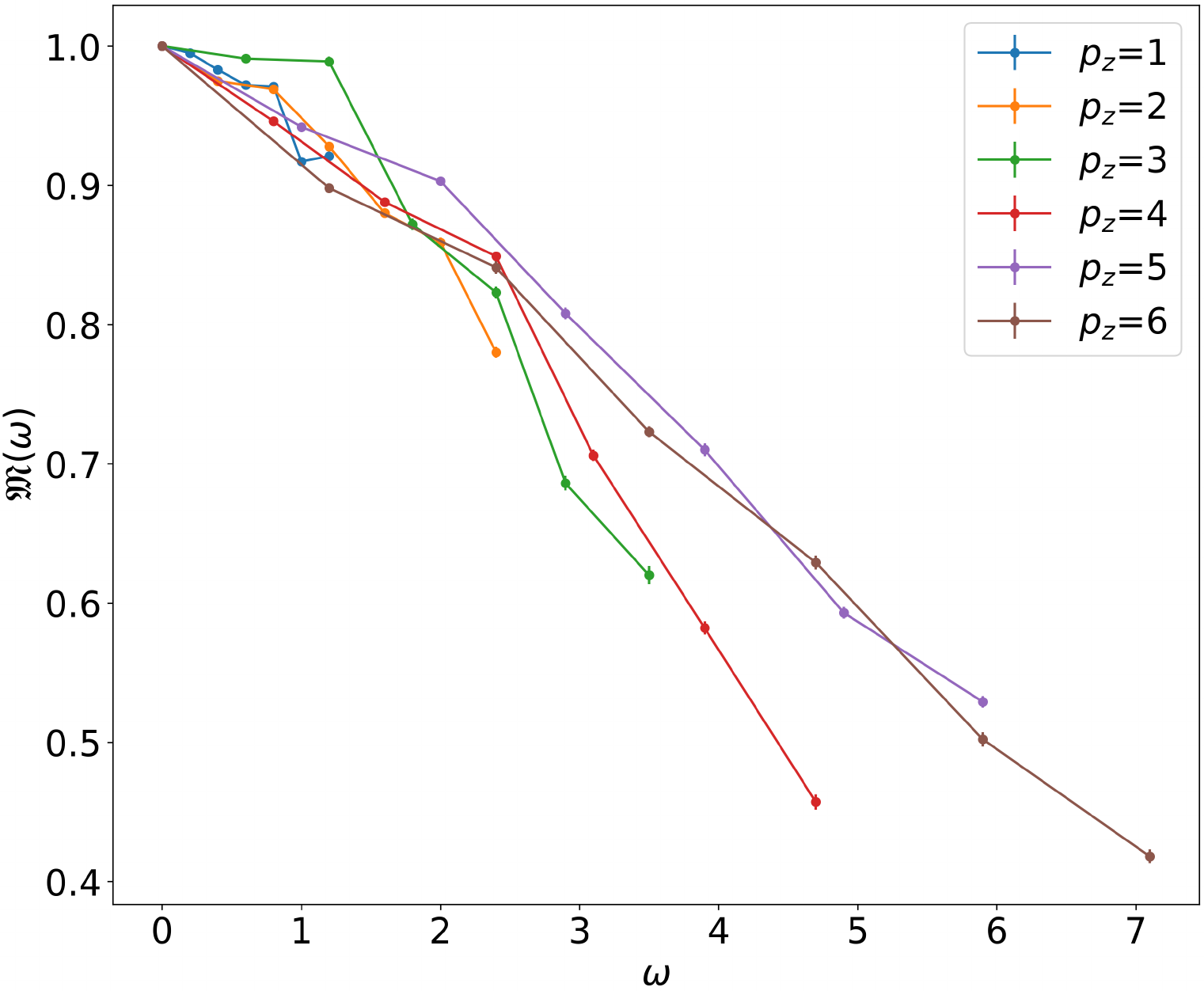}
    \caption{The figure shows the  mean values of $\mathfrak{M}$  from the LQCD calculation against each momenta $p_z$ and different values of $\om$. The mean values are connected with lines. $p_z$ values are denoted by simply $p_z=n$, where $n=1,2,\cdots 6$ according to discrete lattice momenta $p_n = \frac{2\pi n}{La}$. }
\eefs{unpolarized-data}

\befs
\centering
    \includegraphics[width=0.8\textwidth]{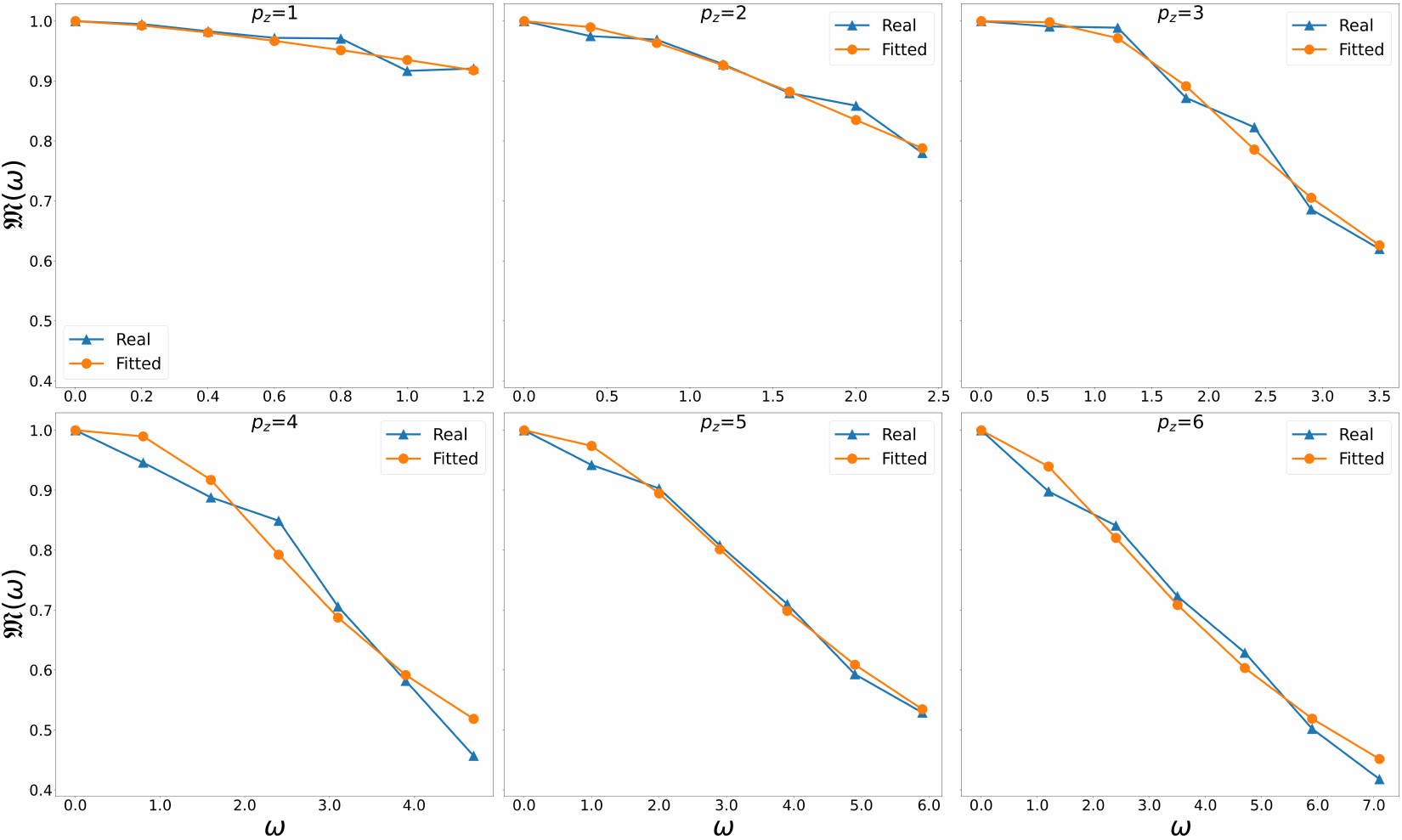}
    \caption{The plot shows the best-fitted mean curve using dose-response curve fitting models. By the blue colored  ``real" data, we refer to the matrix elements obtained from the LQCD calculation, while the orange colored ``estimated" data refers to the ML-fit results.}
\eefs{unpolarized-fit-mean}

\befs
\centering
    \includegraphics[width=0.8\textwidth]{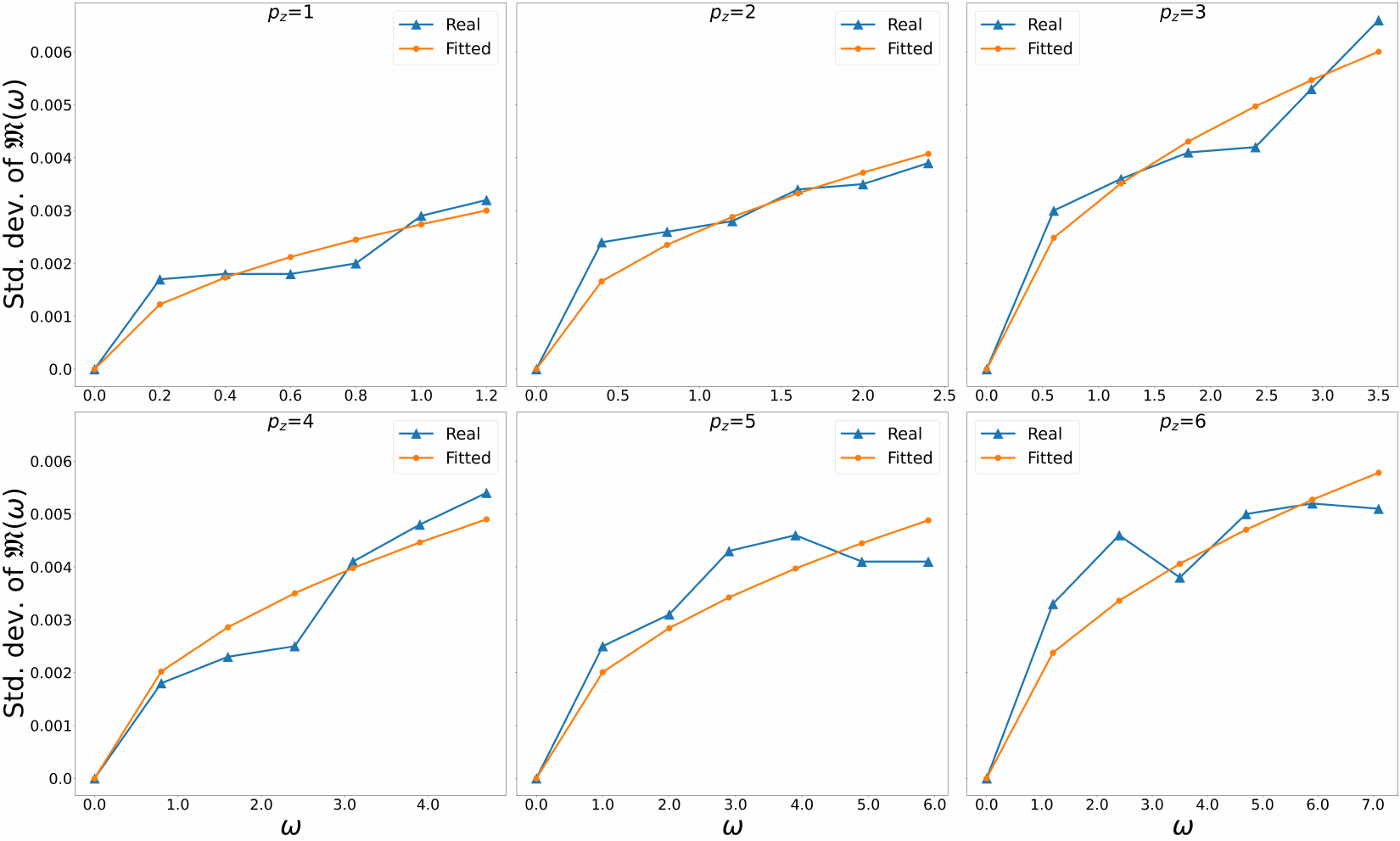}
    \caption{The plot shows the best-fitted standard deviation curve using the gradient descent method. The blue colored ``real" points are obtained from the LQCD calculation, while the orange colored ``estimated" data points are obtained from the ML-fit results.}
\eefs{unpolarized-fit-std}

\noindent
\paragraph{\bf{Curve fitting}:} We used the curve fitting method to find the best-fitted model for mean and standard deviation of $\mathfrak{M}(\om)$ and the fitted model is used to generate mean and standard deviation for new $\om$. We used dose-response~\cite{ritz2015dose} based curve fitting models to fit mean values.  10  models are used to fit a curve, and the best-fitted model is selected using AIC (Akaike Information Criterion) log-likelihood. Fig.~\ref{fig:unpolarized-fit-mean} shows the best-fitted curve for each $p_z$. 

The blue line in Fig.~\ref{fig:unpolarized-fit-std} shows the standard deviation curve for each $p_z$. We tried different non-linear functions (square, cube, and square root) to fit the non-linear standard deviation curves. The square root function shows the best approximation to the real curve. The fitted standard deviation curve for each $p_z$ is represented as a function $f(p_z, \om)=\alpha_z\times\sqrt{\om}$ where $\alpha_z$ is a weight factor. The value of $\alpha_z$ is estimated for the best-fitted curve using the gradient descent method. TABLE~\ref{tab:fit_std_param} shows the estimated $\alpha_z$ for each $p_z$ and Fig.~\ref{fig:unpolarized-fit-std} shows the best fitted standard deviation in orange.

\begin{table}[ht]
    \centering
    \begin{tabular}{|r|c|c|c|c|c|c|}\hline
        $p_z$ & $1$ & $2$ & $3$ & $4$ & $5$ & $6$  \\\hline
        $\alpha_z$ & $0.00274$ & $0.00263$ & $0.00321$ & $0.00226$ & $0.00201$ & $0.00217$  \\\hline
    \end{tabular}
    \caption{The values of the fitted parameter for the matrix elements at each $p_z$.}
    \label{tab:fit_std_param}
\end{table}

The synthetic data is generated using the fitted mean and the standard deviation. The rest of the steps ((a) ML model evaluation and (b) reconstruction and extrapolation of the real data using the best model) are similar to the process we used for the LQCD data associated with the polarized gluon distribution. We applied three ensemble models named gradient-boosted tree (GBT), random forest (RF), and extreme gradient-boosted tree (XGB) (described in Section~\ref{sec:MLMethods}). 

These models' root mean square error (RMSE) in training data are $1.5\times10^{-3}$ for the GBT, $2.2\times10^{-4}$ for the RF, and $5\times10^{-4}$ for the XGB. We find that the RMSE of the GBT is approximately 10 times larger than that of the RF, leading to unacceptable fit results. Therefore, based on the training error and smoothness of the curve, the RF is the best-performing and GBT is the worst-performing model.

\section{Discussion on the ML-generated data and results}\label{sec:MLdisc}
Machine learning is a data-dependent algorithm where the decision process depends on the quality and quantity of the data. Sufficient data is necessary for an ML model to learn the hidden pattern in the data. Due to the lack of these properties in our experimental data, we generated synthetic data to mitigate the gap and achieve the experimental objective. The synthetic data $X$ is randomly generated from a normal distribution $X\sim\mathcal{N}(\mu,\sigma)$ where mean $\mu$ and standard deviation $\sigma$ are predicted using two different machine learning models. The mean is predicted from an ensemble model, and the standard deviation is predicted from a linear regression model. There is uncertainty in each machine learning model to compute mean and standard deviation, which imposes another layer of uncertainty when deriving synthetic data from a normal distribution constructed using that mean and standard deviation. We observed high variations of the error bar for the XGB model in FIG.~\ref{fig:ML-ITDs} compared to the RF and GBT models. This is because of the XGB model's limitation in handling noisy synthetic datasets. This limitation of XGB is reflected in Fig.~\ref{fig:ML-ITDs}. Further investigations with more precise data are needed to quantitatively assess the systematic uncertainties associated with these ML models.

The choice of a proper ML model is critical and mainly depends on the data. The volume and dimension of data exclude the choice of using any deep learning model and leave us to use classical ML models. We tried the linear regression model, but the model did not converge on our data. The support vector regressor with a polynomial kernel also suffers the same issue, but the radial basis function (RBF) kernel cannot outperform ensemble models. Therefore, we stick to ensemble models for our experiment.

The ML model on the standard deviation of the real data for each $\om$ controls the magnitude of the uncertainty. For the polarized data, a linear model fits the polarized data (in Fig.~\ref{fig:std-lr}). The slope of the line indicates the rate of the change of the magnitude of the standard deviation for the change of the $\om$; a steeper slope means the value of the standard deviation will increase a lot for a small increase of the value of $\om$. Alternatively, a gradual slope means a small standard deviation increase for a large increase of the $\om$.  The slope value of Fig.~\ref{fig:std-lr} indicates that $p_z=1$ has a higher slope than $p_z=6$. Therefore, we noticed smaller uncertainty for $p_z=6$ than $p_z=1$ ML-generated dataset.

\subsection{ITD for polarized gluon distribution}
\befs 
\centering

\includegraphics[scale=0.6]{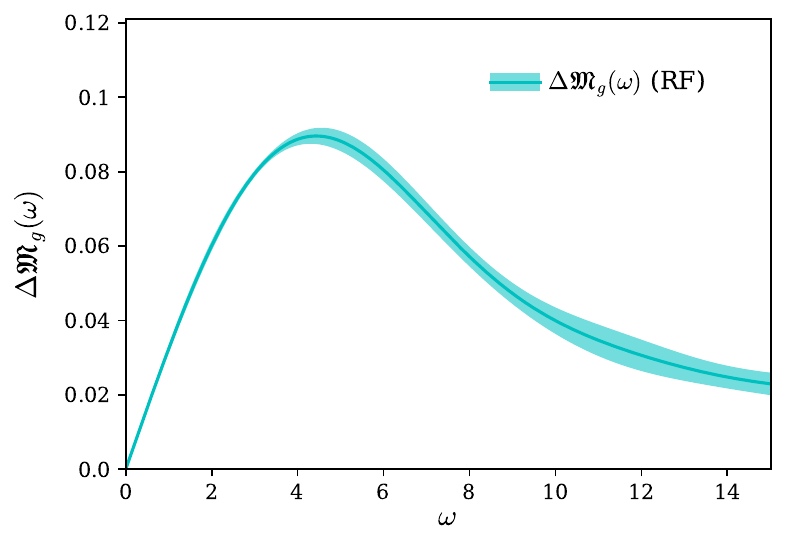}
\includegraphics[scale=0.6]{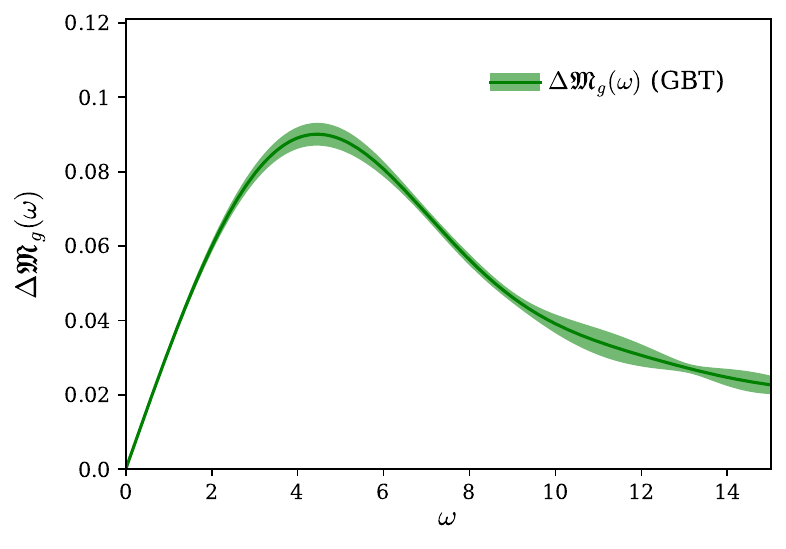}
\includegraphics[scale=0.6]{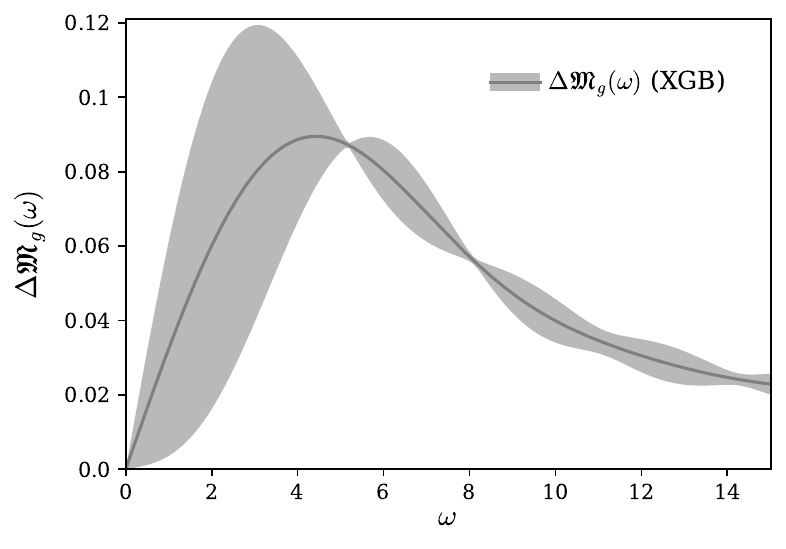}

\caption{\label{fig:ML-ITDs} 
  $\Dl\mathfrak{M}_g(\om)$ after removing contamination term $\frac{m_p^2}{p_z^2} \om \Dl{\mathcal{M}}_{pp}(\om)$ from the LQCD matrix elements $\Dl\mathfrak{M}(\om)$ using  RF, GBT, and XGB machine learning algorithms.}   \eefs{mockdemocn}

In Fig.~\ref{fig:ML-ITDs}, we present the $\Dl \mathfrak{M}_g$ correlation function for the gluon helicity distribution. The contamination-free $\Dl {\mathfrak{M}_g}(\om)$ can now be matched to the light cone  $\Dl{\mathcal I}_g (\om, \mu )$ and  the singlet quark ITD $\Dl{\mathcal{I}}_S (\om, \mu)$ in the $\ms$ scheme using the factorization relation~\cite{Balitsky:2021cwr} up to power corrections,

\begin{eqnarray} \label{eq:matching}
&&\Dl{\mathfrak{M}_g} ( \om ) \langle x \rangle_g(\mu) \!=\!    
\Dl{\mathcal I}_g (\om, \mu )\! -\!  \frac{\alpha_s N_c }{2\pi}\int_0^1 d u\,  \Dl{\mathcal I}_g (u\om, \mu )
    \bigg\{ \ln\bigg(z^2 \mu^2 \frac{e^{2\gamma_E}}{4}\bigg)   \bigg( \bigg[\frac{2u^2} {\bar{u}} + 4u\bar{u}  \bigg]_+ - \bigg(\frac{1}{2}  + \frac{4}{3}  \frac{\langle x_S \rangle{(\mu)}}
 {  \langle x_g \rangle{(\mu)} } \bigg) \delta( \bar{u} ) \bigg) \nn \\
&&+ 4 \bigg[\frac{u+\ln (1-u)}{\bar{u}}\bigg]_+ - \bigg( \frac{1} {\bar u} - \bar{u} \bigg)_+  -\frac{1}{2} \delta(\bar u)       
+2\bar uu \bigg\}- \frac{ \alpha_s C_F}{2\pi}  \int_0^1 d u \,   \Dl{\mathcal{I}}_S (u \om,\mu )  
 \bigg\{\ln \bigg(z^2 \mu^2 \frac{ e^{2 \gamma_E}} {4 } \bigg)  \Dl {\mathcal B}_{gq} (u)+ 2\bar uu    \bigg\}  ,\nn \\
\end{eqnarray}
where $N_c=3$, $\bar{u} \equiv (1 - u)$, $\gamma_E$ is the Euler–Mascheroni constant, $\Dl {\mathcal B}_{gq}= 1-(1-u)^2$,  and   $\langle x \rangle_g(\mu=2~{\rm GeV})$ is chosen to be $0.427$  from~\cite{Alexandrou:2020sml}.  We choose $z=2a$ and  $\mu=2$ GeV in the matching Eq.~\eqref{eq:matching} and ignore the effect of  the singlet-quark contributions (which requires a separate LQCD calculation and is a subject for future investigations). Additionally, varying  values of $z$ or $\mu$ have minimal effects on the matched $\Dl{\mathcal I}_g$ within the current statistical uncertainty due to the large uncertainty of the present LQCD matrix elements~\cite{Khan:2022vot}.

In Fig.~\ref{fig:ML-helITDs}, we present the $\Dl{\mathcal I}_g(\om)$ in the $\ms$ scheme determined from the above-mentioned ML algorithms. We compare the results with the phenomenological ITD constructed from the NNPDF3.1 global fit~\cite{Ball:2017nwa}. As described above, the RF model produces the best result. 

\befs 
\centering

\includegraphics[scale=0.42]{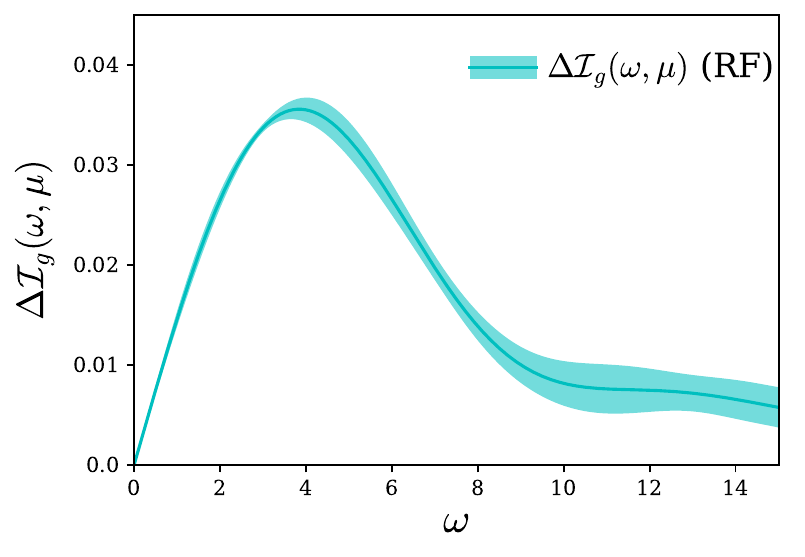}
\includegraphics[scale=0.42]{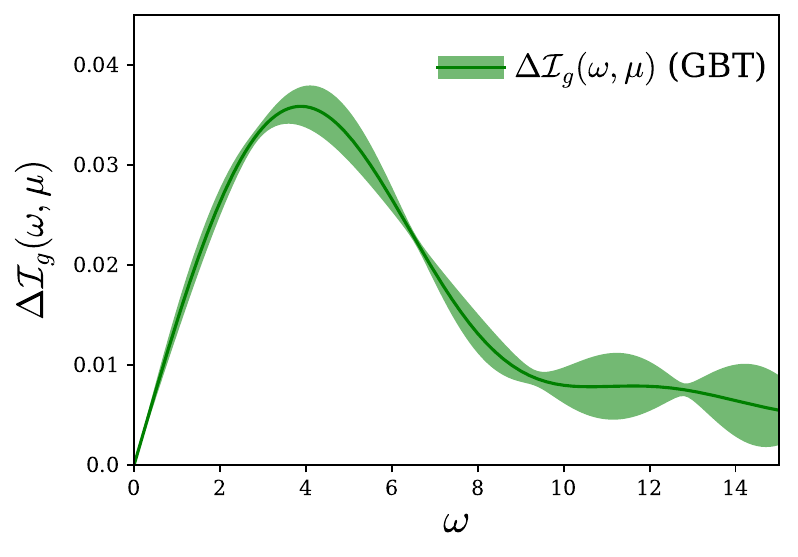}
\includegraphics[scale=0.42]{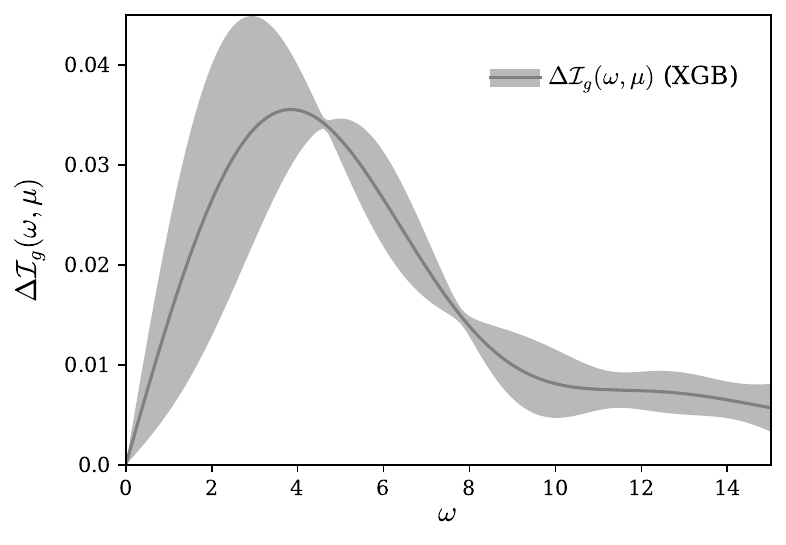}
\includegraphics[scale=0.6]{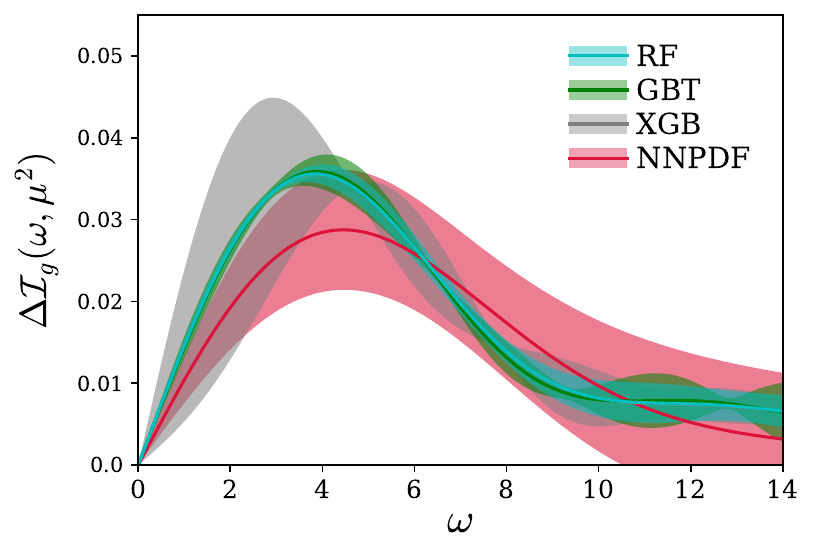}

\caption{\label{fig:ML-helITDs} 
  (Upper panel) Polarized gluon light-cone Ioffe-time distributions from various generative models of machine learning after removing contamination term $\frac{m_p^2}{p_z^2} \om \Dl{\mathcal{M}}_{pp}(\om)$ from  the LQCD matrix elements $\Dl\mathfrak{M}(\om)$ using  RF, GBT, and XGB machine learning algorithm and the implementation of the perturbative matching kernel in the $\ms$ renormalization scheme at 2 GeV. $\Dl\mathfrak{M}_g(\om)$ has been normalized by the gluon momentum fraction $\langle x\rangle_g(\mu)$ before converting to $\Dl\mathcal{I}_g(\om,\mu)$. (Lower panel) We compare our results to phenomenological ITD constructed from the polarized gluon PDFs in NNPDF3.1 global fit~\cite{Ball:2017nwa}.}   \eefs{mockdemocn}

We note the advantages of the ML-generated data as the following:
\begin{itemize}

    \item  The ML cannot fit lattice data beyond $z\geq 0.36$ fm while satisfying the relation~\eqref{eq:master}, exposing the existence of higher twist contaminations in the lattice data at larger spacelike separation $z$.
    \item LQCD calculation of the gluon helicity ITD can have a significant impact in constraining $x\Del g(x)$ and also $\Del G$ in the absence of ample experimental data. The present calculation favors toward ruling out a negative gluon polarization in the nucleon.
    \item The present ML analyses can generate data up to $\om\sim 14$ that is currently inaccessible in any lattice calculations.
    \item  As we will see in Sec.~\ref{sec:PDFs}, the ITD data up to $\om\sim 14$ will be significantly important for reconstructing the PDFs.
    
\end{itemize}

\subsection{ITD for the unpolarized gluon distribution}

Fig.~\ref{fig:unpolarized-fit-m(w)} shows the estimated $\mathfrak{M}(\om)$ and the error bar shows the uncertainty of $\mathfrak{M}(\om)$. As discussed earlier, RF performs the best among the three ML algorithms and GBT performs the worst due to an order of magnitude large RMSE compared to the RF. 

In addition to the description for determining the uncertainty of the unpolarized gluon matrix elements in Sec.~\ref{sec:MLMethods}, one can understand it as follows: We have $351$ experimental (LQCD) data for each $\om (\lesssim 7)$. We compute mean $(\mu_\om)$ and std $(\sigma_\om)$ for each $\om$. Then, for each $\om$, we generate synthetic mean $(\bar{\mu}_\om)$. We randomly pick $351$ points from $N(\bar{\mu}_\om, \sigma_\om)$. This way, up to $\om \approx 7$, we have generated 351 curves for each $p$ and then simply extrapolate those curves and determine the uncertainty at larger $\om$-values. Therefore, similar to the LQCD data, the uncertainties grow with $\om$. Because LQCD data points for  $p_z=1$ are available up to only $\om \sim 1$, the extrapolated $p_z=1$ data has larger extrapolated errors than the $p_z=6$ data. 

In order to determine the ITD in the $\ms$-scheme, we use the one-loop matching relation~\cite{Balitsky:2019krf},

\bea
 \mathfrak{M} (\om, z^2) = \, \frac{\mathcal{I}_g (\om, \mu^2)}{\mathcal{I}_g (0, \mu^2)} - \, \frac{\alpha_s N_c}{2 \pi} \int_0^1 du \; \frac{\mathcal{I}_g (u \om, \mu^2)}{\mathcal{I}_g (0, \mu^2)} 
   \times \Bigg\{ \mathrm{ln}\bigg( \frac{z^2 \mu^2 e^{2 \gamma_E}}{4} \bigg) \; B_{gg} (u) + \; 4 \bigg[ \frac{u + \ln(\bar{u})}{\bar{u}} \bigg]_+ 
  \; \frac{2}{3} \Big[ 1  -u^3 \Big]_+ \Bigg\} \, , \nn \\
 \label{eq:unmatching}
\eea
where we have neglected the quark-gluon mixing. $\mathcal{I}_g (0, \mu^2)$ is the gluon momentum fraction $\langle x\rangle_g(\mu)$ and $B_{gg} (u)$ is the Altarelli-Parisi  kernel. In the left panel of Fig.~\ref{fig:ML-NNPDF-ITDs}, we present the unpolarized gluon ITDs derived from ML-generated lattice data at $p_z=2.05$ GeV and compare these with the phenomenological ITD extracted from the NNPDF3.1~\cite{Ball:2017nwa} in the $\ms$ scheme at $\mu=2$ GeV. A similar comparison is made with the data lattice at $p_z=2.46$ GeV. In contrast to the polarized ITD, where the momentum is very large and thus the $z$-values are very small for $\om = 14$, for the unpolarized ITD with $p_z = 2.46$ GeV, $\om = 14$ corresponds to $z \approx 1$ fm. In  future, with more precise gluonic matrix elements, we aim to isolate the $z$-dependence in the LQCD data and apply ML techniques using both $z$ and $p_z$ as input features. This will allow us to determine the unpolarized gluon ITD at short distances and higher momenta. Unfortunately, current statistics do not permit us to perform such ML analyses.

We note some immediate advantages of the ML-generated data as the following:
\begin{itemize}
    \item The current ML analyses can generate data up to $\om \sim 14$ before the uncertainty bands cross zero. This range of $\om$  is presently inaccessible by any lattice calculations.
    
    \item  The ML-generated ITD up to $\om \sim 14$ will be significantly more important for reconstructing PDFs compared to the ITD limited to $\om \approx 7$ in~\cite{HadStruc:2021wmh}.
\end{itemize}

In the following Sec.~\ref{sec:PDFs}, we extract the PDFs from these ML-generated ITDs and compare them with the phenomenological PDFs.

\befs 
\centering
\includegraphics[scale=0.6]{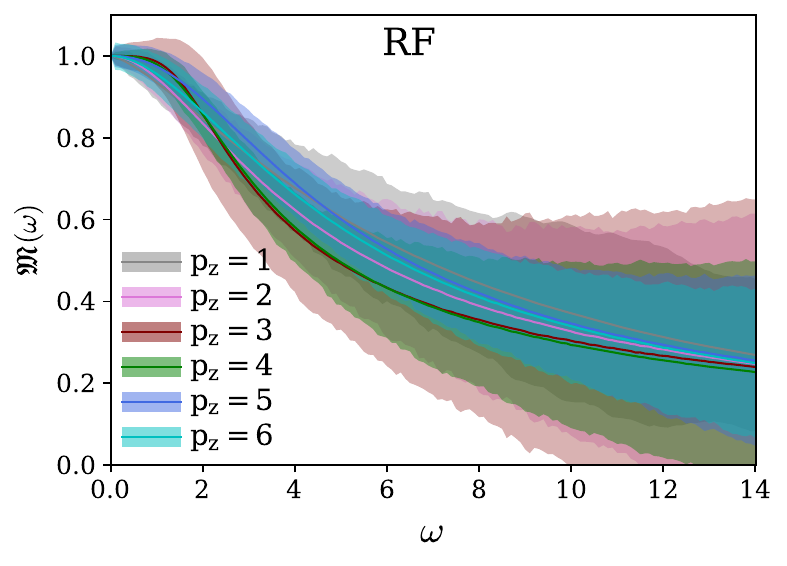}
\includegraphics[scale=0.6]{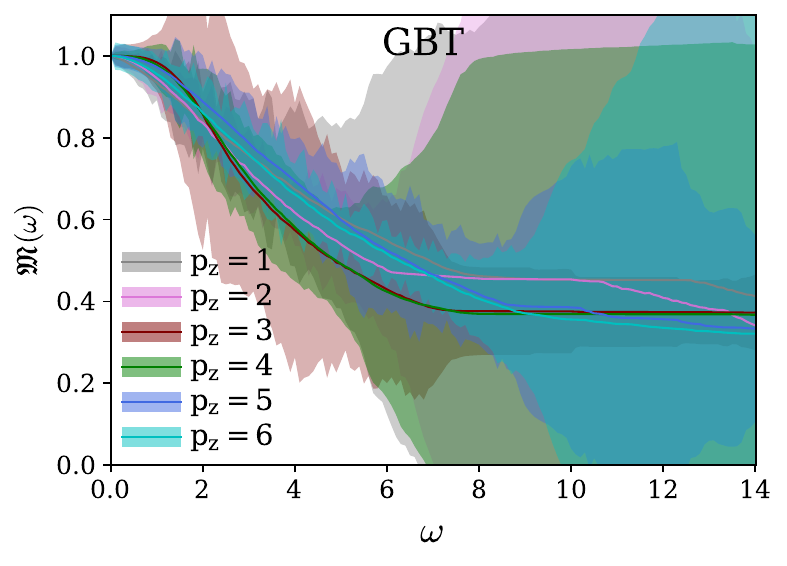}
\includegraphics[scale=0.6]{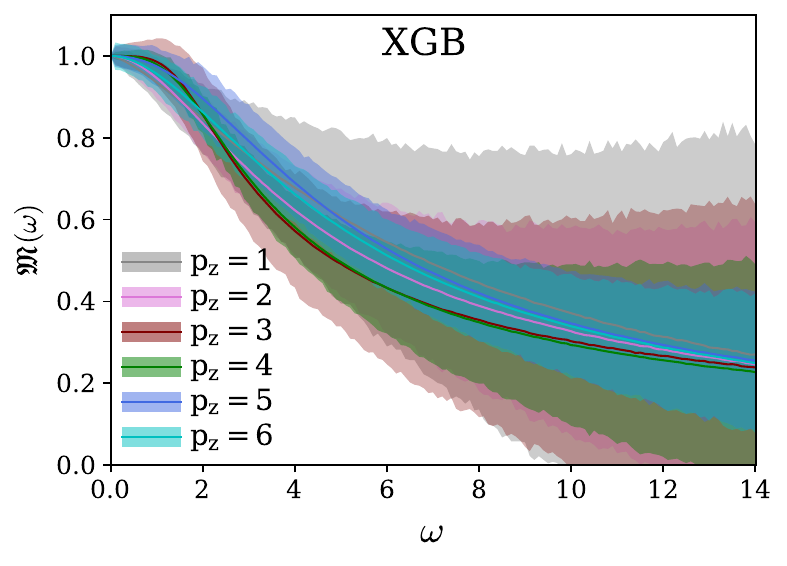}
\caption{\label{fig:unpolarized-fit-m(w)} ML-generation of $\mathfrak{M}(\om)$ for larger values of $\om$ using the three training models.  }   
\eefs{mockdemocn}

\befs 
\centering
\includegraphics[scale=0.6]{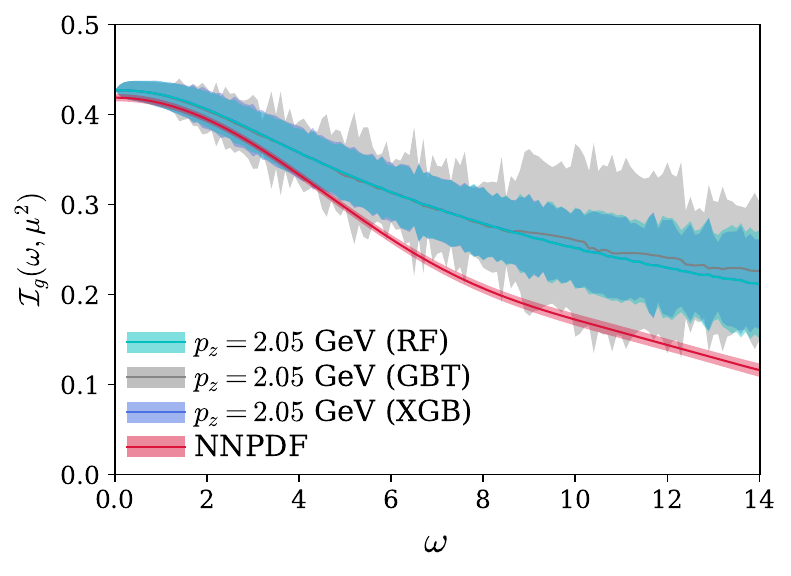}
\includegraphics[scale=0.6]{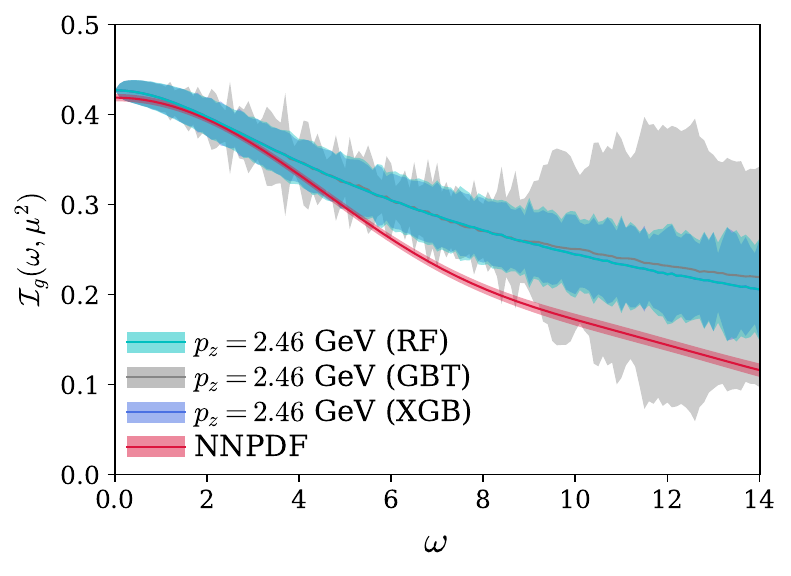}
\caption{\label{fig:ML-NNPDF-ITDs} 
 Unpolarized gluon Ioffe-time distributions from various generative models of machine learning after implementing the perturbative matching kernel in the $\ms$ renormalization scheme at $\mu$ = 2 GeV. We compare our results  with the phenomenological ITD constructed from the unpolarized gluon PDF in NNPDF3.1 global fit~\cite{Ball:2017nwa}. }    
\eefs{mockdemocn}


\section{Neural network reconstruction of PDFs}\label{sec:PDFs}

To extract the polarized gluon PDF $x\Dl g(x,\mu)$ from the ITDs presented in the previous section, we need to solve the following inverse problem~\cite{Braun:1994jq}:
\bea \label{eq:NNpol}
\Dl \mathcal{I}_g(\om,\mu) = \int_0^1 \dd x~x\Dl g(x,\mu)\sin(x\om).
\eea
Similarly, the unpolarized gluon PDF can be determined by solving the following inverse problem:
\bea \label{eq:NNunpol}
\mathcal{I}_g(\om,\mu) = \int_0^1 \dd x~xg(x,\mu)\cos(x\om). 
\eea
We determine the $x\Dl g(x)$ and $xg(x)$ distributions in Eqs.~\eqref{eq:NNpol} and~\eqref{eq:NNunpol} using the similar neural network (NN) architecture used in~\cite{Khan:2022vot}. Instead of directly parametrizing the distribution function by a NN, we take a prefit function for the polarized gluon PDF from one of the parametrizations in~\cite{Sufian:2020wcv},
\begin{align}
    \Dl h_0(x) = xg_{+0}(x) - xg_{-0}(x),
\end{align}
where
\begin{align}\label{eq:prefitpol}
    xg_{+0} &= 20.2\, x^{0.025} (1-x)^{4.97}  (1 - 2.91 \sqrt{x} + 2.47 x)[1 - 0.87(1-x)],\\
    xg_{-0} &= 20.2\, x^{0.025} (1-x)^{6.97}  (1 - 2.91 \sqrt{x} + 2.47 x)[1 - 0.87(1-x)].
\end{align}
We note that starting from a point closer to the solution can significantly improve the efficiency of the fitting. Though mathematically equivalent to a direct parametrization, it can accelerate the convergence of the fitting to a smooth function, while the result is not sensitive to any particular reasonable choice of the prefit function $\Dl h_0(x)$. Then the polarized gluon distribution $x\Dl g(x)$ is parametrized as $\Dl h_0(x)$  multiplied by an NN. The architecture of the  NN contains an input layer for $x$ values, and the output layer  neurons  for $x\Dl g(x)$, which is transformed to ITD via~\eqref{eq:NNpol} and compared with the LQCD data. Three hidden layers, containing 32, 32, and 8 neurons respectively, are inserted between the input layer and the output layer. They are densely connected to the corresponding former layers and activated with the rectified linear unit function. The output layer is densely connected to the last hidden layer and activated with the sigmoid function, which is renormalized and shifted to return values between $-10$ to $10$, allowing both positive and negative distributions within a large reasonable range for the  polarized gluon distribution. 

The fitting procedure is to minimize the loss function defined as the $\chi^2$ between the NN predictions and the data points. In each fit, we randomly select 80\% of the data points generated by the ML algorithms in the previous sections for $\om \leq 10,12,14$ and leave the remaining 20\% of the data points in the validation sample  to evaluate the performance of the model on unseen data. To keep the possibility of finding multiple minima, the initial parameters of the NN are randomly generated. The loss value of the full dataset is monitored during the training. It generally decreases at the beginning and starts to increase when overfitting happens, with small fluctuations from epoch to epoch all the time.  To prevent the accidental selection of small loss value points, some early epochs are eliminated. We stop the training process when there is no improvement in the total loss value for $3000$ epochs and revert to the best result obtained. The result from the epoch with the least total loss function is saved. The results always converge to the same region. In addition, the partition of the training sample and the validation sample was not fixed from time to time, and the result does not show dependence on  the partition within the uncertainties inheriting from the  LQCD  data. As has been investigated in~\cite{Khan:2022vot}, the systematic uncertainty from the choice of the NN is negligible and the uncertainty is dominated by the lattice data. One may perform the analysis using an NN with an additional hidden layer, {\it i.e.} an NN with four hidden layers containing 32, 32, 32, and 8 neurons respectively as more hidden layers mean more flexible parametrization of the function. We find almost no noticeable effects on the resulting distribution function with the uncertainty. One can also modify the activation functions of the layers, {\it e.g.} the exponential linear unit function and the hyperbolic tangent function. Once again, similar outcomes of the PDFs are obtained. Since these modifications of the NN have already been investigated and numerically demonstrated in~\cite{Khan:2022vot}, we do not repeat the calculations here. Rather, we concentrate on the PDF reconstructions for different ranges of $\om\in [10,14]$ which have the largest effects on the resulting PDFs. 

Based on the outcomes in Sec.~\ref{sec:ML1}, we consider the ITDs from the two best fits, from RF and GBT. In Fig.~\ref{fig:polPDF-NN}, we present $x\Dl g(x)$ reconstructed from  the ITD data up to $\om_{\rm max}=10,12,14$ and compare  them  with that presented in~\cite{Ball:2017nwa}. 

\befs 
\centering
\includegraphics[scale=0.6]{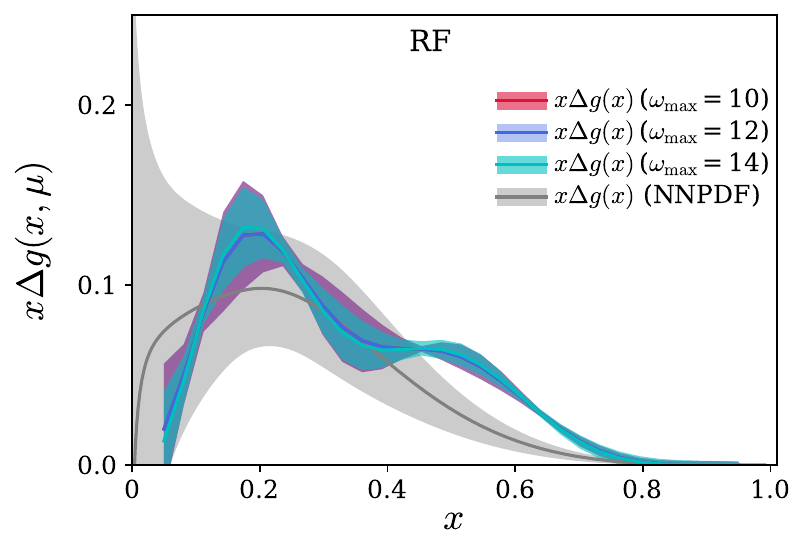}
\includegraphics[scale=0.6]{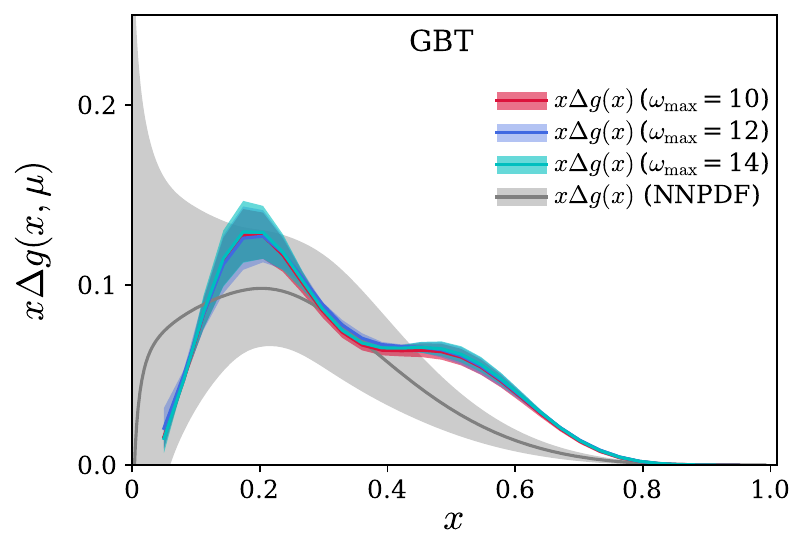}
\caption{\label{fig:polPDF-NN} 
Neural network reconstruction of the polarized gluon PDF $x\Dl g(x)$ from the ITDs generated with RF and GBT machine learning algorithms. The neural network reconstructions of the PDFs have been done for three different values of $\om_{\rm max}=10,12,14$ of the ITD, $\Dl \mathcal{I}_g(\om,\mu)$. The grey band is obtained from the NNPDF global analysis in~\cite{Ball:2017nwa}. }    
\eefs{mockdemocn}

From Fig.~\ref{fig:polPDF-NN}, we can observe that the reconstructed PDFs are very different from the prefit function~\eqref{eq:prefitpol} which supports the discussion above. Since the data only cover a finite range of $\om$ and the PDF at small $x$ is more sensitive to large $\om$ extrapolation  of the ITD, the NN reconstruction of the PDF breaks down when $x$ is small and we only present the distribution  for $x > 0.05$. On the other hand, the distribution function at large $x$ is small and thus requires higher numerical precision. For efficiency, we only fit the distribution up to $x=0.95$ and the endpoint at $x=1$ is fixed to zero by the parametrization. Therefore, we constrain the fit in the range of $[0.05 \leq x \leq 0.95]$.  When using a neural network to extract the  $x\Delta g(x)$, it attempts to match the ITD data within a limited range, inferring the missing information beyond this range based on the features learned from the available data. The inverse problem we are solving is analogous to a Fourier transform, where $x\Dl g(x)$ can be thought of as a distribution over frequencies. If the range of $\om$ is restricted, it is similar to a truncation of the Fourier series, which will introduce oscillations in the reconstructed function, as can be observed from the extracted distribution around $x\sim 0.5$. These oscillations are due to the inability to perfectly resolve $x\Dl g(x)$ from a finite set of data. More discussions about the significance of the resulting PDFs will be discussed in the following Sec.~\ref{sec:PDFdisc}.

Similarly, for the unpolarized gluon distribution, we take the prefit function from a fit in~\cite{Sufian:2020wcv},
 \begin{align}\label{eq:unpolprefit}
      h_0(x) = xg_{+0}(x) + xg_{-0}(x),
 \end{align}
and parametrized $xg(x)$ as $h_0(x)$ multiplied by an NN. The architecture of the NN is the same as the one for the polarized distribution above, but the last layer is normalized and shifted to return values between $0$ to $10$, a  range large enough  to cover any reasonable results of  the PDFs. Our parametrization has assumed that the unpolarized gluon distribution is positive definite. As discussed above, the GBT algorithm fails to fit the LQCD data. We, therefore, use the ITDs from  the RF and XGB  algorithms for the reconstruction of the PDFs and present the results in Fig.~\ref{fig:latt-NN} for the ITD data up to $\om_{\rm max}=10,12,14$.

As pointed out in Sec.~\ref{sec:extrapol}, the restricted functional form of  the PDFs can be biased and severely underestimate uncertainties. Therefore, we avoid such PDF reconstructions. However, if  the ITD data is not available  in a sufficiently large $\om$-domain, the PDF reconstruction can be challenging and produce large errors. In this regard, the ML-generated ITD outside the LQCD accessible $\om$ range can be very useful for the reconstruction of  the PDFs using the NN. The corresponding results are shown in Fig.~\ref{fig:largex}. Here are some important remarks: One has to be careful as the reconstructed PDFs in the region $x \lesssim 0.2$ have smaller uncertainties for the ITD with $\om_{\rm max}=7$ compared to the ITD with $\om_{\rm max}=14$. This indicates the possibility that the ML-generated ITD data are  not highly sensitive to the $xg(x)$ distribution in the $x \lesssim 0.2$ domain. One therefore must be cautious and not consider the $xg(x)$ distribution highly reliable in the $x \lesssim 0.2$ domain.

In contrast, as shown in the PDF reconstruction in Fig.~\ref{fig:largex}, it is immediately evident that the  ITD with $\om_{\rm max}=14$  is much more effective for PDF reconstruction compared to the  ITD with $\om_{\rm max}=7$  in constraining uncertainties in the mid- and large-$x$ regions of the constructed PDF. Specifically, the LQCD data is expected to be most sensitive to the  PDF's mid-$x$ region, where significantly smaller uncertainties can be achieved with the  ITD with $\om_{\rm max}=14$ compared to the ITD with $\om_{\rm max}=7$. Another significant achievement is that the ML-generated ITD can reduce uncertainties in the mid to large $x \sim 0.9$ region. A precise estimate of $xg(x)$ at large $x$ is crucial for generating accurate predictions of both signal and background in searches for new massive particles at the LHC~\cite{Nocera:2017zge}. Since there is a difference between the  ITD from the NNPDF  dataset  and  ITD from the LQCD calculation at an unphysical pion mass and a coarse lattice spacing (see Fig.~\ref{fig:ML-NNPDF-ITDs}), we expect the reconstructed PDF to differ from the NNPDF. The major advantage highlighted here is that with  the LQCD and  the ML-generated data up to $\om = 14$, the impact on constraining the $xg(x)$ distribution at mid to large $x$ can be significant.

\befs 
\centering
\includegraphics[scale=0.6]{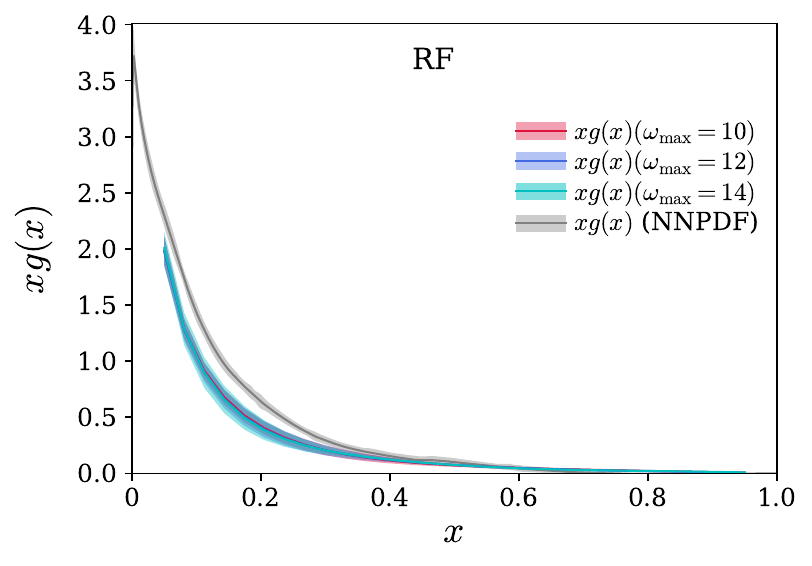}
\includegraphics[scale=0.6]{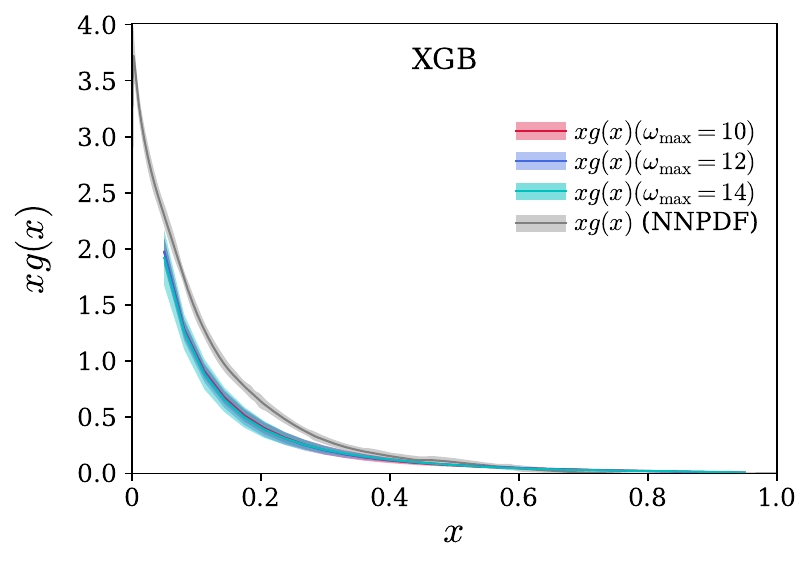}
\caption{\label{fig:latt-NN} 
Neural network reconstruction of  the unpolarized gluon PDF $x g(x)$ from  the ITDs generated with RF and GBT machine learning algorithms. The neural network reconstructions of  the PDFs have been done for three different values of $\om_{\rm max}=10,12,14$ of the ITD, $\mathcal{I}_g(\om,\mu)$. The grey band is obtained from the NNPDF global analysis in~\cite{Ball:2017nwa}.}    
\eefs{mockdemocn}

\befs 
\centering
\includegraphics[scale=0.7]{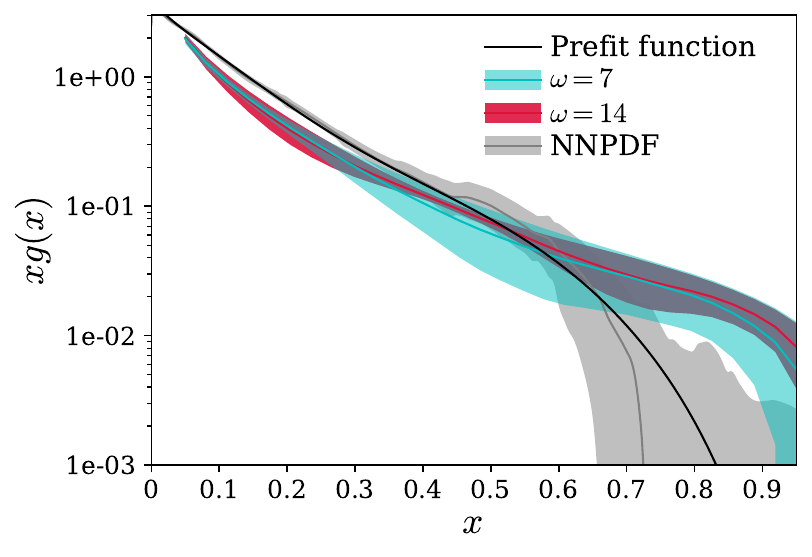}
\caption{\label{fig:largex} Neural network reconstruction of the unpolarized gluon PDF using the RF-generated data for $\om_{\rm max}=7$ and $14$. We compare our results with the gluon PDFs extracted from NNPDF3.1 global fit~\cite{Ball:2017nwa}. The prefit function~\eqref{eq:unpolprefit} is also plotted.}    
\eefs{mockdemocn}
 
\section{Impacts of the polarized and unpolarized gluon distribution results}\label{sec:PDFdisc}

In the first LQCD determination of the gluon helicity PDF~\cite{Khan:2022vot}, it was predicted that the gluon helicity PDF  is positive in the mid- to large-$x$ region within uncertainty. This contrasts with two global analyses~\cite{Zhou:2022wzm,Whitehill:2022mpq}, which reported that both positive and negative solutions for  $x \Delta g(x)$  were equally capable of describing the experimental data. Subsequently, a similar conclusion from~\cite{Khan:2022vot} was also found in~\cite{deFlorian:2024utd} with the fundamental requirement that  the physical cross-sections must not be negative. Following~\cite{deFlorian:2024utd}, the analyses in~\cite{Zhou:2022wzm,Whitehill:2022mpq} were revisited in~\cite{Hunt-Smith:2024khs}, and a similar observation of the positivity of $x \Delta g(x)$ was found. In the present work, using the physics-informed generative ML algorithms, we have solidified our previous results in~\cite{Khan:2022vot} with higher precision and up to a larger value of $\om \sim 14$. While caution is needed regarding the sign of $x \Delta g(x)$ in the small-$x$ region, where LQCD data is not  sensitive  to $x \Delta g(x)$, the PDFs determined in Fig.~\ref{fig:polPDF-NN} show a positive distribution in the interval $x \in [0.05, 0.95]$. A recent phenomenological calculation~\cite{Chakrabarti:2023djs} has also determined $x\Dl g(x)$ to be positive. In fact, this finding is also consistent with the LQCD calculation~\cite{Yang:2016plb} at the physical pion mass, continuum, and infinite volume limits. The LQCD calculation in~\cite{Yang:2016plb} obtained the $x$-integrated value of $\Dl G = 0.251(47)(16)$ using a local matrix element~\cite{Ji:2013fga}. 

Additionally, it has been found that the ML fits cannot describe the LQCD data well for  $z>0.36$ fm. This possibly indicates that across all different $p_z$, the data points for $z > 0.36$ fm do not satisfy Eq.~\eqref{eq:master}, which is imposed to constrain the outcome from the ML algorithms. This also indicates that the LQCD data have significantly higher twist contributions for $z > 0.36$ fm. Therefore,  the LQCD data at $z > 0.36$ fm are not suitable for extracting the leading-twist dominated $\Dl \mathfrak{M}_g(\om)$. With future precise gluonic matrix elements, it remains an important subject to determine up to which $z_{\rm max}$ LQCD data is dominated by the leading-twist contribution, and generative ML can serve as a useful tool for these studies.

For the unpolarized gluon distribution, the results presented in the previous Secs.~\ref{sec:ML1} and~\ref{sec:PDFs} show a clear advantage of the application of the generative ML on the LQCD data. For the gluonic matrix elements, it might not be possible to achieve precise ITD up to $\om=14$ in the near future just from LQCD calculations. Our  calculation also suggests avoiding model-dependent parameterizations of the PDFs  to prevent bias and underestimation of  uncertainties. Even with a large range of  $0 \lesssim \om \lesssim 14$, we find that $xg(x)$ can be reliably determined only in the $0.2 \lesssim x \lesssim 0.95$ interval. On the other hand, applying generative ML to  the LQCD data to generate ITDs up to $\om\sim 14$ can provide a significant advantage in addressing the inverse problem of determining $xg(x)$.

As noted in the previous section, the $xg(x)$ falls off much more slowly than the NNPDF result at large $x$. This may be due to the LQCD calculation being performed at a heavier pion mass and a coarse lattice spacing. Additionally, we observe that because of the large uncertainty in the LQCD matrix elements, the ML algorithms could not isolate any $z$-dependence from the LQCD matrix elements and used data up to $z = 0.56$ fm (unlike the polarized data, where Eq.~\eqref{eq:master} required the use of the LQCD data up to $z = 0.36$ fm). As discussed in Sec.~\ref{sec:MLdisc}, the large $z$ data used in the ML-generated ITD might be affected by  the higher-twist contaminations, leading to a slower fall-off of the PDF. However, as mentioned earlier, the goal of this work is to demonstrate how ML can open an avenue to constrain $xg(x)$ distribution in the mid-to-large $x$ region using  LQCD calculations. Future precise LQCD matrix elements will enable isolating the $z$-dependence as an input feature in the ML analyses to obtain $xg(x)$ in the physical and continuum limits. In addition, physics-informed ML application on the LQCD data can have a significant impact on extracting  the gluon generalized parton distributions (GPDs). These prospects have been discussed in~\cite{Kriesten:2021sqc,Allaire:2023fgp,Liuti:2024zkc}.

\section{Conlusion and outlook}\label{sec:con}
We have presented analyses demonstrating how the synergy between LQCD and generative ML can effectively facilitate the determination of  the unpolarized and  the polarized gluon distributions,  that  current LQCD calculations alone cannot achieve. We utilized three different ML algorithms to identify the most suitable ones for representing the LQCD data and generating  the Ioffe-time distributions up to $\om \sim 14$, thereby significantly  alleviating the inverse problem of determining PDFs from LQCD calculations. Our work shows a positive gluon helicity PDF in the $x\in [0.05, 0.95]$ interval. This suggests a positive contribution of  the glue spin to the proton spin budget within this $x$-window. 

We have extracted the PDFs from the ML-generated data using  neural network  and demonstrated the limitations of relying on model-dependent and strongly constrained functional forms. This approach helps avoid the potential underestimation of uncertainties in both the fitted  LQCD data and the resulting PDFs. Finally, we have shown how the  combination of LQCD and ML can have a significant impact on constraining the unpolarized gluon distribution in the mid- to large-$x$ regions.

This work also demonstrates that LQCD nonlocal spatial matrix elements for the gluon helciity distribution beyond $0.36$ fm may contain significant higher-twist contamination, which can impact  the PDF determination and should be avoided to ensure the applicability of factorizing  the LQCD matrix elements into perturbative Wilson coefficients and nonperturbative PDFs with controllable power corrections. To address this issue, one can train the ML algorithms to learn from short-distance correlations at $z\lesssim 0.36$ fm, thereby minimizing possible contamination from higher-twist effects $\mathcal{O}(\Lambda_{\rm QCD}^2z^2)$ at large spatial distances for a successful reconstruction of  the PDFs.  With future precise lattice QCD matrix elements, we expect, ML can estimate the correlations between all different $z$ and $p_z$ data points and paramterize the  leading power
corrections, such as $\mathcal{O}(z^2\Lambda^2)$ in the short distance factorization
and $\mathcal{O}(\Lambda^2/(xp_z)^2),~\mathcal{O}(\Lambda^2/((1-x)p_z)^2)$ in the LaMET. Looking forward, the demonstrated interface between LQCD and ML holds promising potential for investigating  the higher-twist contributions at the level of LQCD correlation functions when precise data are available using $z$, $p_z$, and $\om$ as input features in the ML, an area that remains largely unexplored.

Currently, the error bands are primarily due to statistical uncertainties, but with more precise data in the future, systematic uncertainties can be  quantified. Given the existing challenges in extracting  the polarized and  the unpolarized gluon PDFs, it is crucial that one can use lattice QCD calculations to complement global analyses, potentially opening a new avenue for understanding the role of gluons in the nucleon in the mid- to large-$x$ regions, where the PDFs are less constrained by the experimental data.

\section{Acknowledgement}

R.S.S. thanks Peter Boyle and Swagato Mukherjee who provided insight,
valuable suggestions, and expertise that greatly assisted
this research. T.I. is supported by the U.S.
Department of Energy (DOE) under award DE-SC0012704, SciDAC-5 LAB 22-2580, and also
Laboratory Directed Research and Development (LDRD No. $23-051$) of BNL and RIKEN-BNL Research Center. M.K. is supported by the National Technology Engineering Solutions of Sandia DE-NA$0003525$ with the U.S. Department of Energy (DOE). T.L. is supported by the National Natural Science Foundation of China under Grants No. $12175117$ and No. $12321005$, and Shandong Province Natural Science Foundation Grant No. ZFJH$202303$.  J.S. is supported by the U.S. Department of Energy through Contract No. DE-SC0012704 and by Laboratory Directed Research and Development (LDRD) funds from Brookhaven Science Associates. R.S.S. is supported by the Laboratory Directed Research and Development (LDRD No. $23-051$) of BNL and RIKEN-BNL
Research Center.

\bibliography{References.bib}


\end{document}